\documentclass[11pt, preprint]{aastex}
\usepackage{amssymb}
\usepackage{natbib}
\usepackage{epsfig}

\newcommand       \Angstrom     {\,{\rm \AA}}

\newcommand       \um           {\mu{\rm m}}
\newcommand       \mum          {\,{\rm \mu m}}
\newcommand       \Ks           {{\rm K_{S}}}
\newcommand       \J            {{\rm J}}
\newcommand       \HH           {{\rm H}}
\newcommand       \K            {{\rm K}}
\newcommand       \simali       {\,{\sim}}
\newcommand       \magni        {\,{\rm mag}}

\newcommand       \Rv           {{R_V}}
\newcommand       \AV           {{A_V}}
\newcommand       \Av           {{A_V}}
\newcommand       \AB           {{A_B}}

\newcommand       \EBV          {{E({\rm B-V})}}
\newcommand       \StoN          {{\rm S/N}}
\newcommand       \simlt        {\lesssim}
\newcommand       \simgt        {\gtrsim}
\newcommand       \gtsim        {\gtrsim}

\countdef\decade=200
\advance\decade by \year
\countdef\hours=201
\advance\hours by \time
\divide\hours by 60
\countdef\mins=202
\advance\mins by \hours
\multiply\mins by 60
\multiply\hours by 100
\countdef\miltime=203
\advance\miltime by \hours
\advance\miltime by \time
\advance\miltime by -\mins
\def\today{\number\decade.\number\month.\number\day.\number\miltime}

\shorttitle{Mid-IR Extinction in the LMC}
\shortauthors{Gao et al.\ }

\begin{document}

\title{
The Mid-Infrared Extinction Law in the Large Magellanic Cloud
\\{\small DRAFT: \today ~~}
       }

\author{Jian Gao\altaffilmark{1,2},
B.~W. Jiang\altaffilmark{1,2},
Aigen Li\altaffilmark{2},
and M.~Y. Xue\altaffilmark{1}}

\altaffiltext{1}{Department of Astronomy,
                 Beijing Normal University,
                 Beijing 100875, China;
                 {\sf jiangao@bnu.edu.cn, bjiang@bnu.edu.cn}
                 }
\altaffiltext{2}{Department of Physics and Astronomy,
                 University of Missouri,
                 Columbia, MO 65211, USA;
                 {\sf lia@missouri.edu}
                 }

\begin{abstract}
Based on the photometric data
from the \emph{Spitzer}/SAGE survey
and with red giants as the extinction tracers,
the mid-infrared (MIR) extinction laws
in the Large Magellanic Cloud (LMC)
are derived for the first time
in the form of $A_\lambda/A_{\Ks}$,
the extinction in the four IRAC bands
(i.e., [3.6], [4.5], [5.8] and [8.0]$\mum$)
relative to the 2MASS $\Ks$ band at 2.16$\mum$.
We obtain the near-infrared (NIR)
extinction coefficient to be
$E(\J-\HH)/E(\HH-\Ks) \approx 1.29\pm0.04$
and $E(\J-\Ks)/E(\HH-\Ks) \approx 1.94\pm0.04$.
The wavelength dependence of the MIR extinction
$A_{\lambda}/A_\Ks$ in the LMC varies
from one sightline to another.
The overall mean MIR extinction is
$A_{[3.6]}/A_\Ks\approx0.72\pm0.03$,
$A_{[4.5]}/A_\Ks\approx0.94\pm0.03$,
$A_{[5.8]}/A_\Ks\approx0.58\pm0.04$, and
$A_{[8.0]}/A_\Ks\approx0.62\pm0.05$.
Except for the extinction in the IRAC [4.5] band
which may be contaminated
by the 4.6$\mum$ CO gas absorption of red giants
(which are used to trace the LMC extinction),
the extinction in the other three IRAC bands show a flat curve,
close to the Milky Way $R_V=5.5$ model extinction curve
(where $R_V$ is the optical total-to-selective extinction ratio).
The possible systematic bias caused by the correlated
uncertainties of $\Ks-\lambda$ and $\J-\Ks$ is explored
in terms of Monte-Carlo simulations. It is found that this
could lead to an overestimation of $A_{\lambda}/A_\Ks$ in
the MIR.
\end{abstract}



\section{Introduction}\label{sec:intro}
The Large Magellanic Cloud (LMC) is a low-metalicity
irregular dwarf galaxy and a satellite of the Milky Way (MW).
Since the metallicity of the LMC (which is only $\simali$1/4
of that of the MW; \citealt{Russell92}) is similar to that of
galaxies at red shifts $z\sim1$ \citep{Dobashi08},
it offers opportunities to study the dust properties
in distant low-metallicity extragalactic environments
by studying the extinction properties of the LMC.

The wavelength dependence of interstellar extinction
-- ``interstellar extinction law (or curve)'' --
is one of the primary sources of information
about the interstellar grain population \citep{Draine03}.
In the MW, the interstellar extinction laws in the ultraviolet (UV)
and visual wavelength ranges vary from sightline to sightline,
and can be characterized by the optical total-to-selective
extinction ratio $\Rv \equiv \AV /\EBV$ \citep{Cardelli89},
where $\EBV=\AB-\AV$ is the interstellar reddening,
$\AV$ is the extinction at the visual
($V$; $\lambda_V\approx 5500\Angstrom$) band,
and $\AB$ is the extinction at the blue
($B$; $\lambda_B\approx 4400\Angstrom$) band.
The regional variations of the UV/visual extinction curves
(i.e., the variations of the $\Rv$ values) reflect the variations
in dust size: larger $\Rv$ values indicate the predominance of
larger grains \citep{Draine03}.
However, the infrared (IR) interstellar extinction laws of the MW,
which also vary from one sightline to another, cannot be simply
represented by the single $\Rv$ parameter.
Many recent studies show that there does not exist
a ``universal'' near-infrared (NIR) extinction law
for the MW \citep{Fitzpatrick09, Gao09}.
Moreover, the observationally-determined mid-infrared (MIR)
extinction law shows a flat curve
while classical dust models for
the diffuse interstellar medium (ISM)
predict a much steeper curve,
with a pronounced minimum at
$\simali$7$\mum$ \citep{Draine89}.\footnote{%
  In this work by ``NIR'' we mean $1\mum < \lambda < 3\mum$
  and by ``MIR'' we mean $3\mum < \lambda < 8\mum$.
  }

\begin{figure}
\centering
\includegraphics[angle=0,width=5in]{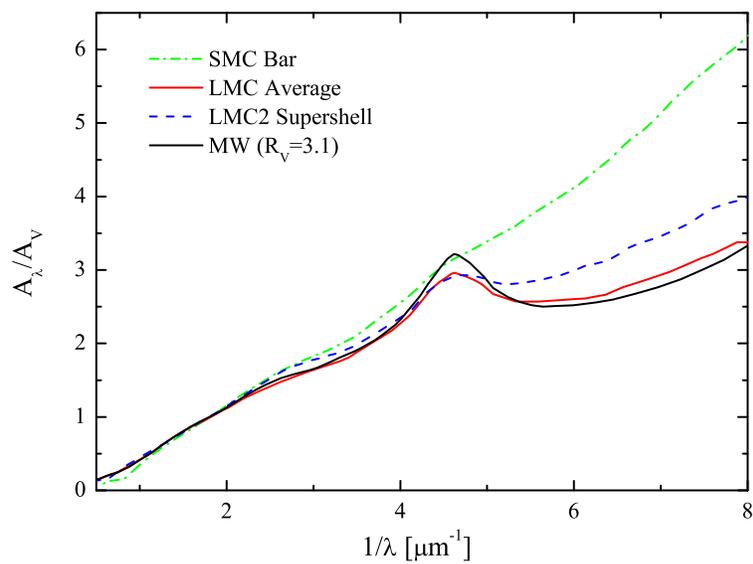}
\caption{\footnotesize
         \label{fig:mwlmcsmc}
         Comparison of the MW (black solid),
         LMC (red solid), and SMC (green dot-dashed)
         extinction curves.
         Also shown is the LMC2 supershell
         extinction curve (blue dashed).
         The MW extinction curve is calculated
         from the CCM parameterization with $\Rv=3.1$.
         The LMC and SMC extinction curves
         are taken from \citet{Gordon03}.
         }
\end{figure}

Due to its low metallicity, the dust quantity (relative to H)
in the LMC is expected to be lower than that of the MW because
there is less raw material (i.e., heavy elements) available for
making the dust. The (relative) lack of the dust-making raw material
could prevent the dust in the LMC from growing and hence the dust
in the LMC may be smaller than the MW dust.
Furthermore, the star-formation activity
in the LMC could destroy the dust.
Therefore, one would naturally expect the dust size distribution
and extinction curve in the LMC to differ from that of the MW.
As illustrated in Figure~\ref{fig:mwlmcsmc},
the Galactic interstellar extinction curve rises from
the NIR to the near-UV with a broad absorption bump
at about $\lambda^{-1}\approx4.6\um^{-1}$
($\lambda \approx 2175 \Angstrom)$,
and continues rising steeply into
the far-UV \citep{Draine03}.
In the Small Magellanic Cloud (SMC), the extinction curves
of most sightlines display a nearly linear steep rise with
$\lambda^{-1}$ and lack the 2175$\Angstrom$ hump \citep{Lequeux82,Prevot84}.
The LMC extinction curve is intermediate between
that of the MW and that of the SMC:
compared to the Galactic extinction curve,
the LMC extinction curve is characterized by
a {\sl weaker} 2175$\Angstrom$ hump
and a {\sl stronger} far-UV rise \citep{Nandy81,Koornneef81,Gordon03}.
Strong regional variations in extinction properties
have also been found in the LMC \citep{Clayton85,Fitzpatrick85,Fitzpatrick86,Gordon98,Misselt99,Gordon03}:
the sightlines toward the stars inside or near
the supergiant shell, LMC\,2, which lies on
the southeast side of the 30 Doradus star-forming region,
have a very weak 2175$\Angstrom$ hump \citep{Misselt99},
while the extinction curves for the sightlines toward the stars
which are $>$\,500\,pc away from the 30 Doradus region are
closer to the Galactic extinction curve.
\citet{Gordon03} estimated $\Rv\approx2.76$ for the LMC\,2 supershell
and $\Rv\approx3.41$ for the LMC as a whole.
The sightlines outside 30 Doradus have $\Rv \approx 3.2$ \citep{Fitzpatrick86,Gordon98,Misselt99}.
\citet{Koornneef82} had already noticed that
the average value of $\Rv$ in the LMC ($\Rv\approx3.01$)
is close (within 10\%) to that of the MW ($\Rv\approx3.1)$.
We note that the CCM parameterization controlled by
the single $\Rv$-parameter is not valid for the LMC
extinction curve \citep{Gordon03},
i.e., even the LMC and the MW are close in $\Rv$,
their extinction curves differ appreciably.

\citet{Koornneef82} derived the NIR extinction law of the LMC
based on the NIR photometry at the $\J$, $\HH$, and $\K$ bands
of early type supergiants.
He obtained $E(\rm V-\J)/E(\rm B-\rm V)\approx2.26\pm0.23$,
$E(\rm V-\HH)/E(\rm B-\rm V)\approx2.58\pm0.27$,
and $E(\rm V-\K)/E(\rm B-\rm V)\approx2.91\pm0.31$,
and argued that the NIR extinction law of the LMC
is very similar to that of the MW.\footnote{%
   For the average Galactic extinction,
        the corresponding color ratios are
        $E(\rm V-\J)/E(\rm B-\rm V)\approx 2.255$,
        $E(\rm V-\HH)/E(\rm B-\rm V)\approx 2.588$,
        and $E(\rm V-\K)/E(\rm B-\rm V)\approx 2.780$ \citep{Koornneef82},
        or $E(\rm V-\J)/E(\rm B-\rm V)\approx 2.22\pm0.02$,
        $E(\rm V-\HH)/E(\rm B-\rm V)\approx 2.55\pm0.03$,
        and $E(\rm V-\K)/E(\rm B-\rm V)\approx 2.744\pm0.024$ \citep{Rieke85}.
    }
Their results correspond to $E(\J-\HH)/E(\HH-\Ks)\approx1.06$
when converted to the 2MASS photmetric system \citep{Imara07}.
Because of this, the Galactic NIR extinction law is sometimes adopted
for the LMC (e.g., \citealt{Cioni00,Imara07}).
Using the near-infrared color-excess (NICE) method \citep{Lada94},
\citet{Imara07} calculated the NIR extinction coefficients of the LMC
to be $\Av/E(\HH-\Ks)\approx20.83\pm0.52$
and $\Av/E(\J-\HH)\approx17.30\pm0.46$,
corresponding to $E(\J-\HH)/E(\HH-\Ks)\approx1.20\pm0.04$,
which is roughly consistent with that of \citet{Koornneef82}.
We note that although the LMC NIR extinction law is
commonly assumed to be universal,
\citet{Gordon03} showed that
it also differs between the LMC\,2 supershell
and the LMC average (see their Table~4).

So far, little efforts have been put into
the MIR extinction properties of the LMC
due to the paucity of MIR data.
Nevertheless, the MIR extinction is not negligible:
the LMC optical extinction $\Av\simlt5.0\magni$
\citep{Imara07,Dobashi08} would imply an appreciable amount
of MIR extinction $A_{[3.6]}\simlt0.35\magni$
if we take the Galactic ratio of
$A_{[3.6]}/\Av\sim 0.07$ \citep{Gao09}.
With the advent of sensitive IR space facilities
(e.g., {\sl ISO} and {\sl Spitzer}),
the LMC has been mapped in the MIR wavelength bands
with high accuracy and this makes the exploration of
the LMC MIR extinction possible.
In this work, based on the \emph{Spitzer}/SAGE database,
we probe the MIR extinction of the LMC
and its regional variations.
In \S2 we briefly describe the SAGE data used in this work.
\S3 presents the method adopted to derive the extinction,
including the Galactic foreground extinction correction,
the LMC sightline selection, and the selection of
extinction tracers. \S4 reports the derived extinction law
and the mean extinction of the LMC.
Finally, we summarize our results in \S5.

\section{Data}\label{DATA}
\begin{figure}
\centering
\includegraphics[angle=0,width=3.6in]{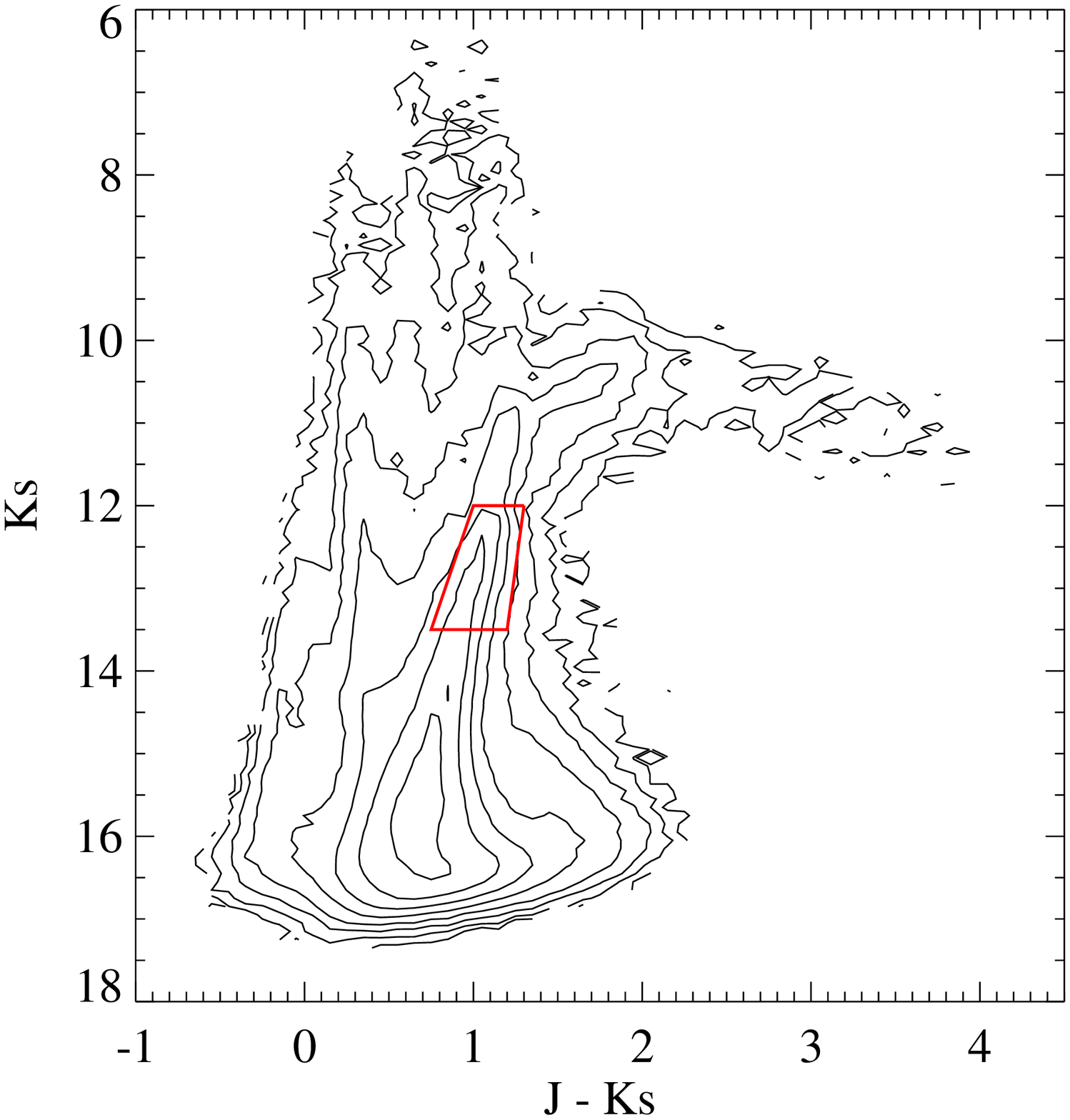}
\hspace{-0.83in}
\includegraphics[angle=0,width=3.6in]{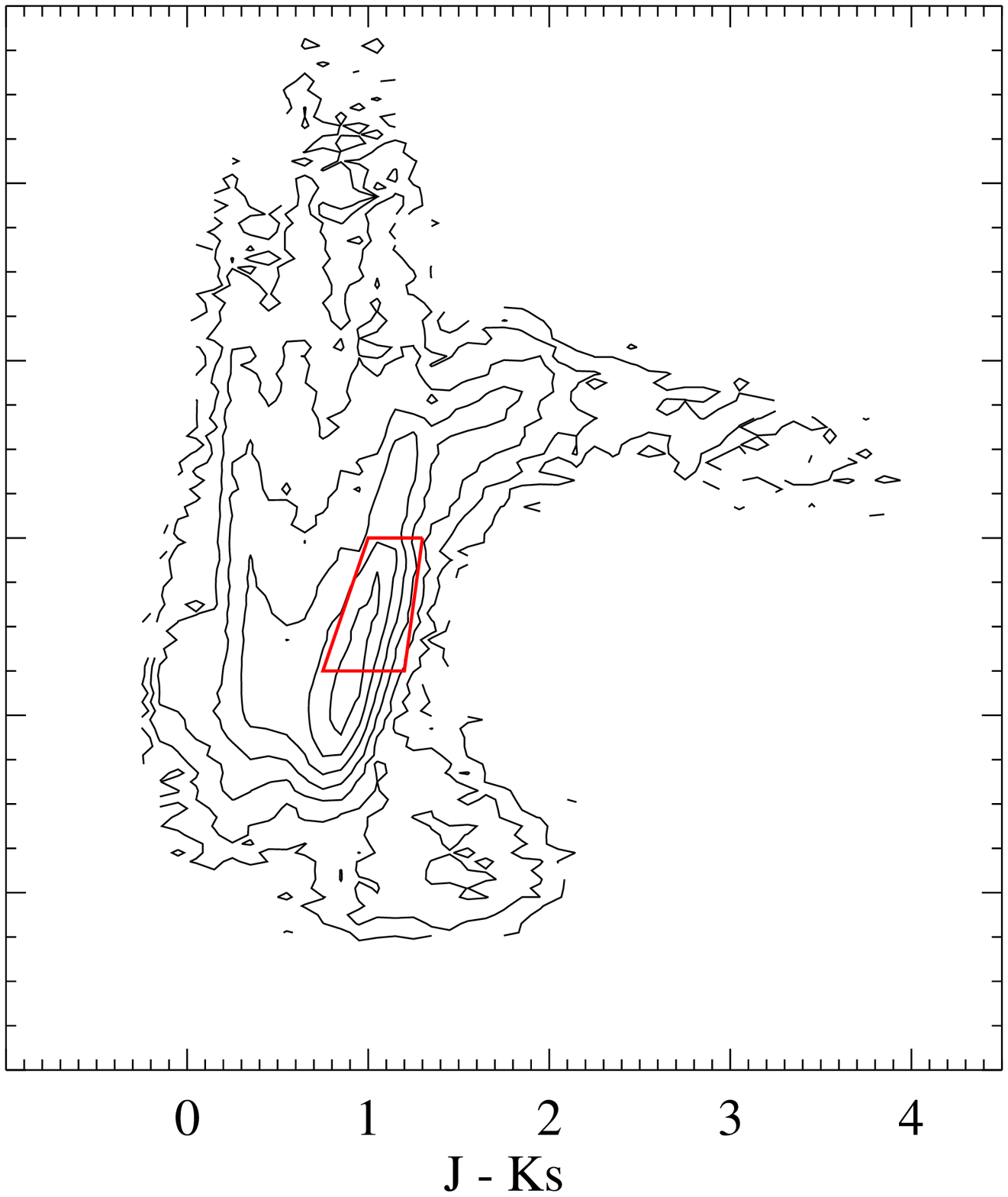}
\caption{\footnotesize
         \label{fig:cmd}
          The color-magnitude diagram (CMD) of the LMC.
          Left panel shows all the sources in the catalog of
          SAGELMCcatalogIRAC, while the right panel only shows
          the sources with S/N\,$\simgt$\,5 at all three 2MASS bands
          and four IRAC bands.
          The red trapezium shows the region of RGBs,
          which are candidates for extinction tracers.
          }
\end{figure}

The data used in this work were obtained
through the \emph{Spitzer}/SAGE Legacy Program,
entitled ``\emph{Spitzer} Survey of the Large Magellanic Cloud:
Surveying the Agents of a Galaxy's Evolution'' \citep{Meixner06}.
The SAGE legacy program mapped the LMC at two different epochs
(epochs 1 and 2) separated by three months,
using the IRAC ([3.6], [4.5], [5.8], and [8.0]$\um$)
and MIPS ([24], [70], and [160]$\um$) instruments
on board the \emph{Spitzer} Space Telescope.
The SAGE Points Source Catalog, named SAGELMCcatalogIRAC,
released in September 2009,
combined both epochs' data bandmerged with 2MASS and 6X2MASS
all-sky data \citep{Cutri03,Cutri04}.
The SAGELMCcatalogIRAC catalog is the SAGE catalog with
the highest quality, providing $\simali$6.4 million point sources
at four IRAC bands and three 2MASS or 6X2MASS bands.
Faint limits are 18.1, 17.5, 15.3, and 14.2$\magni$
for IRAC [3.6], [4.5], [5.8], and [8.0]$\mum$, respectively.
The SAGE data also provide the catalogs,
such as SAGELMCcatalogMIPS24, SAGELMCcatalogMIPS70
and SAGELMCcatalogMIPS160, which include the sources
observed by \emph{Spitzer}/MIPS
at [24], [70], and [160]$\um$ bands.
However, only IRAC data are used in this work
for probing the LMC MIR extinction.
We only select the sources with a signal-to-noise ratio of
S/N\,$\simgt$\,5 at all three 2MASS bands and four IRAC bands
(see \S\ref{tracer} for details on sample selection).
Table \ref{tab:obserr}
shows the mean photometric errors for each band
in the SAGELMCcatalogIRAC
with $\StoN\simgt5$.
Because of the high sensitivity of {\it Spitzer}/IRAC,
there are $\approx293935$ sources in SAGELMCcatalogIRAC
with S/N\,$\simgt$\,5 at all seven bands (see Figure~\ref{fig:cmd}).
Additionally, \citet{Meixner06} estimated that the Galactic foreground stars
and background galaxies contribute roughly 18\% and 12\% of
the SAGE catalog sources.

\section{Methods and Tracers} \label{method}
\subsection{Foreground Extinction}
Toward the LMC, many studies have showed that
the mean foreground Galactic reddening is
$E(\rm B-\rm V)\sim0.06\magni$ \citep{Bessell91,Oestreicher95,Staveley-Smith03,Imara07}.
\citet{Dobashi08} estimated the average Galactic extinction
across the LMC at the visual band to be $\Av\sim0.2\magni$,
implying $E(\rm B-\rm V)\sim0.06\magni$
with the Galactic average value of $\Rv=3.1$.\footnote{%
   \citet{Israel86} and \citet{Imara07} found that $E(\rm B-\rm V)$
   may vary from $\simali$0.01$\magni$ to $\simali$0.14$\magni$
   toward different parts of the LMC.
   \citet{Schwering91} found the foreground reddening
   $E(\rm B-\rm V)\sim0.05\magni$ toward 30 Doradus,
   while the inner reddening in this area exceeds 0.14$\magni$.
   }

We correct for the foreground extinction at each 2MASS band,
taking the foreground visual extinction
to be $\Av=0.2\magni$ \citep{Dobashi08}
for all the LMC sources. To see how much the extinction will change
using different foreground extinction,
we re-calculated the extinction toward CO-186 (30 Doradus).\footnote{%
            The foreground visual extinction
            would be $\Av\sim0.43\magni$
            if we take $E(B-V)=0.14\magni$,
            the upper limit of the foreground reddening
            \citep{Israel86,Imara07}.
            With $\Av=0.43$ and $\Rv=3.1$,
            the foreground extinction at the IR bands
            are $A_{\J}\approx0.121$, $A_{\HH}\approx0.075$,
            $A_{\Ks}\approx0.048$, $A_{[3.6]}\approx0.030$,
            $A_{[4.5]}\approx0.027$, $A_{[5.8]}\approx0.024$,
            and $A_{[8.0]}\approx0.026\magni$.
            Taking these foreground extinction quantities
            to correct the observed magnitude,
            there is little change to
            the derived LMC NIR extinction
            (compared to that with $\Av=0.2\magni$):
            the difference is $\simlt$2.1\% and $\simlt$2.3\%
            for $E(\J-\HH)/E(\HH-\Ks)$
            and $E(\J-\Ks)/E(\HH-\Ks)$, respectively,
            while the LMC MIR extinction $E(\Ks-\lambda)/E(\J-\Ks)$
            at [3.6]$\mum$, [5.8]$\mum$, and [8.0]$\mum$
            would decrease by $\simali$16\%, 24\%, and 33\%,
            respectively.
            If neglecting the foreground reddening
            (i.e., taking $E(B-V)=0$), the derived NIR extinction
            also changes little ($\simlt$1.3\%, $\simlt$0.7\%),
            while the MIR extinction at [3.6]$\mum$, [5.8]$\mum$,
            and [8.0]$\mum$ would increase by $\simali$0.8\%, 6.4\%,
            and 9.2\%, respectively.
            The extinction at the [4.5]$\mum$ band is more
            complicated\textbf{ (see \S\ref{ice45})}.
            }
We take the wavelength-dependence of the foreground extinction
to be that of the $\Rv=3.1$ Galactic average extinction law
\citep{Rieke85}, i.e.,
$A_{\J}/\AV \approx0.282$,
$A_{\HH}/\Av \approx0.175$, and
$A_{\Ks}/\Av \approx0.112$.
The mean foreground extinction at each 2MASS band
are thus $A_{\J}\approx0.056$, $A_{\HH}\approx0.035$,
and $A_{\Ks}\approx0.022\magni$.
The mean foreground extinction derived from the extinction map of
\citet{Schlegel98} are $A_{\J}\approx0.055$,
$A_{\HH}\approx0.035$,
and $A_{\Ks}\approx0.021\magni$ \citep{Imara07}.
For the four IRAC bands, the foreground extinction
is also corrected by applying the Galactic MIR extinction of
$A_{[3.6]}/A_{\Ks} \approx0.63$,
$A_{[4.5]}/A_{\Ks} \approx0.57$,
$A_{[5.8]}/A_{\Ks} \approx0.49$,
and $A_{[8.0]}/A_{\Ks} \approx0.55$,
which were obtained for 131 \emph{Spitzer}/GLIPMSE fields
along the Galactic plane within $|l| < 65^{\rm o}$
with red giants and red clump giants (RCGs) as tracers \citep{Gao09}.
Therefore, the mean foreground extinction at the four IRAC bands are
$A_{[3.6]}\approx0.014$,
$A_{[4.5]}\approx0.013$,
$A_{[5.8]} \approx0.011$,
and $A_{[8.0]} \approx0.012$$\magni$, respectively.

\subsection{Selection of Regions} \label{sightline}
As mentioned in \S1 the LMC UV/visual extinction exhibits
substantial regional variation, particularly among sightlines
toward stars in and outside the 30 Doradus region.
Little is known about the MIR extinction of
the LMC and its variation toward different sightlines,
until the \emph{Spitzer}/SAGE database offers us
the unique opportunity.
As the MIR extinction is generally small,
to probe the MIR extinction efficiently
we favor the LMC regions with large $\Av$
(after all, only these highly obscured regions
need to be corrected for IR extinction).

Previous studies have estimated that the mean LMC reddening
varies from $E(\rm B-\rm V)\approx0.10\magni$ to
$E(\rm B-\rm V)\approx0.16\magni$
(see Table~2 in \citealt{Imara07}).
\citet{Imara07} constructed the visual extinction ($\Av$) map
and found the mean value of $\Av$ is $\simali$0.38$\magni$.
\citet{Zaritsky04} constructed extinction maps for
two stellar populations in the central 64 $\rm deg^{2}$
area of the LMC, and derived the average extinction
of $\AV\approx 0.43\magni$ and $\AV\approx 0.55\magni$
for the cold and hot populations, respectively.
\citet{Dobashi08} developed a new method
by using the color of the \emph{X} percentile reddest stars
and derived a new $\Av$ map of the LMC,
which is similar to the integrated intensity
of the CO emission as observed by
the NANTEN telescope \citep{Fukui08}.

All these extinction maps showed that the maximum $\Av$
of the LMC is close to $\simali$5$\magni$.
Since $A_{\Ks}/\AV\sim0.1$ (the extinction at the $\Ks$ band
relative to $\Av$), for $\Av=1\magni$
$A_{\Ks}$ is only $\simali$0.1$\magni$.
The mean measurement uncertainty of the 2MASS survey
is $\sim$0.109\,mag for the $\Ks$ magnitude (SNR\,=\,1).
Therefore, it is necessary to only consider the regions
with $\Av\simgt1\magni$.
In addition, the LMC areas containing special structures
(such as bar or molecular ridge)
and some special interstellar environments
also need to be considered in order to probe
the variation of MIR extinction
among different interstellar environments.

\begin{figure}
\epsscale{.50} \plotone{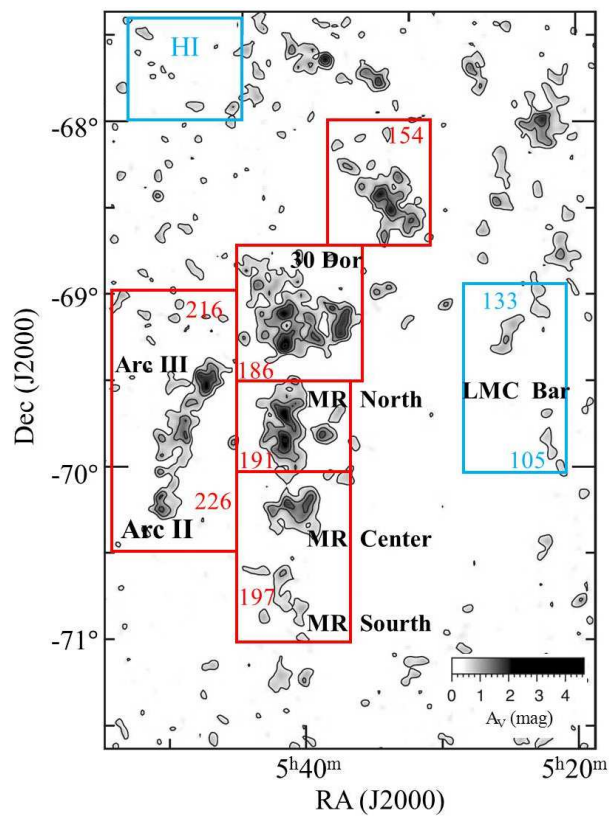}
\centering
\caption{\footnotesize
          \label{fig:fields}
          The seven LMC regions selected for studying the
          LMC MIR extinction. The red boxes show the
          five regions selected to probe the MIR extinction of the LMC.
          The two blue boxes show the regions in the LMC bar
          and in the HI area, where $\Av$ is smaller than 1$\magni$.
          The extinction map is taken from \citet{Dobashi08}
          (see their Figure~6).
           }
\end{figure}

\citet{Sakon06} investigated the mid- to far-IR emission
of the LMC based on the \emph{COBE}/DIRBE
and \emph{IRAS} Sky Survey Atlas data.
In their Figure~12, they illustrated the local structures
in the molecular ridge (MR) and the CO arc in the LMC.
These structures are associated with the CO molecular clouds,
as well as the highly obscured regions
in the extinction map derived by \citet{Dobashi08}.
Combining \citet{Dobashi08}'s extinction map
and \citet{Sakon06}'s Figure~12
with the NANTEN $^{12}$CO map \citep{Fukui08},
we select five regions with large $\Av$
in the LMC molecular clouds and MR and arc structures
(see the red boxes in Figure~\ref{fig:fields};
in the following we will call these five regions
``the first five regions'').
Although the average $\Av$ is smaller than 1$\magni$
or even much smaller, the regions in the LMC bar and HI area
(marked as blue boxes in Figure~\ref{fig:fields})
are also considered for comparison.
All these seven regions are listed in
Table~\ref{tab:field} labeled with
the structure name and the molecular cloud number.

\subsection{Color-Excess Method\label{sec:cem}}
\begin{figure}
\centering
\includegraphics[angle=0,width=7.2in]{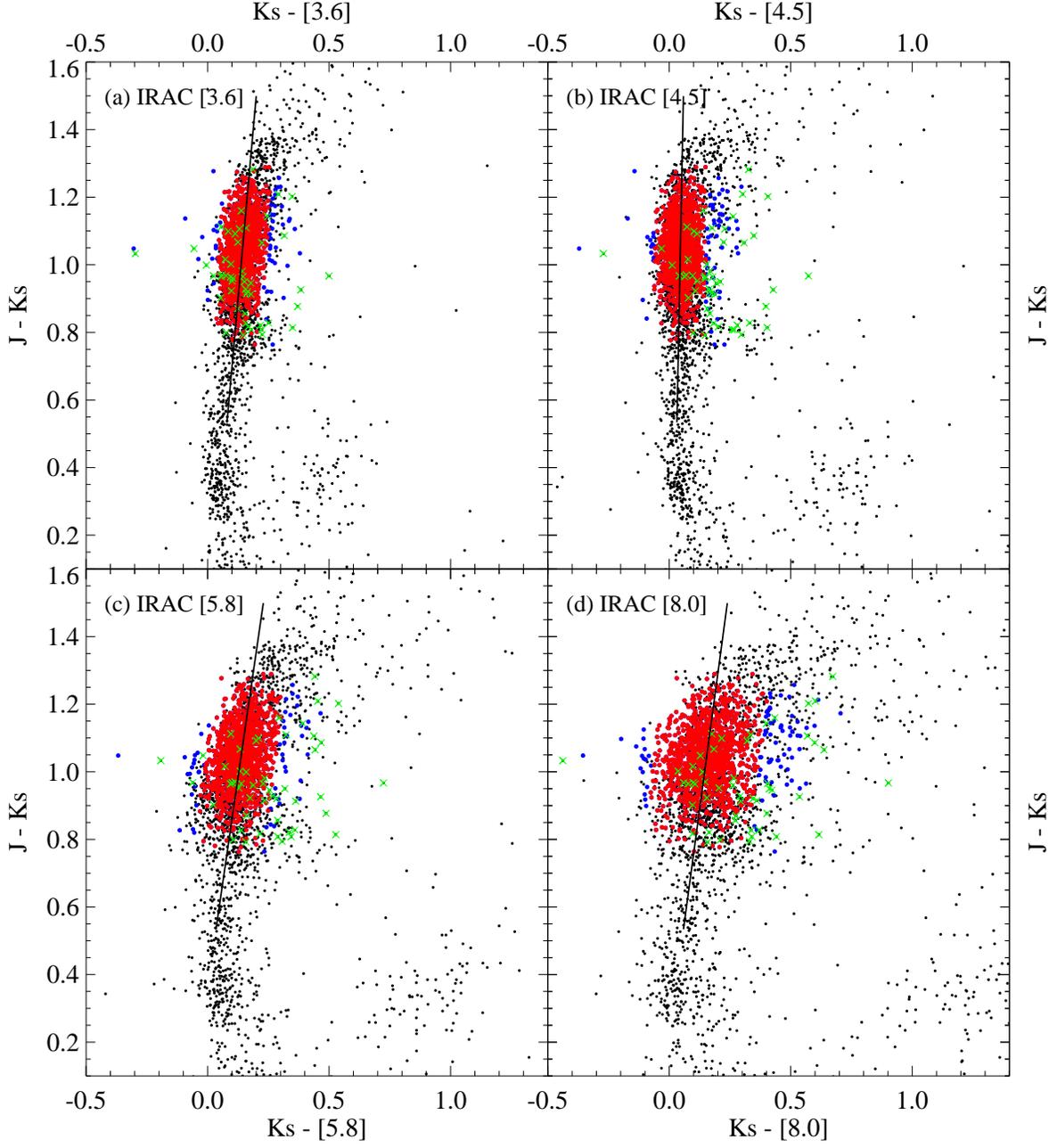}
\caption{\footnotesize
         Sample of fitting for the 30 Doradus region (CO-186).
         The slopes of the black lines are $1/k_{\lambda}$
         (see eq.\,\ref{eq_ext} for $k_{\lambda}$, which can be used to
         derive $A_{\lambda}/A_{\rm Ks}$).
         Red and blue dots show all RGB sources selected
         in the 30 Doradus region (CO-186)
         using the method described in \S\ref{tracer},
         while green crosses show the sources in the \emph{Region} E
         with $[3.6]-[4.5]> 0$ or $[5.8]-[8.0] > 0.5$.
         Only red dots are the RGB sources used in the final fitting
         after applying the $3\sigma$ principle.
         Black dots show all the sources with S/N\,$\simgt$\,5
         in all seven (2MASS and IRAC) bands in this region.
\label{fig:fit}}
\end{figure}

The determination of dust extinction is most commonly made
by comparing the flux densities of obscured and unobscured
pair stars of the same spectral type \citep{Li12}.
In this work we adopt the ``color-excess'' method
to obtain the extinction.
The ``color-excess'' method is widely applied to photometric
data and can probe deeper than the spectrum-pair method.
For doing this, a group of sources which have essentially
the same intrinsic color indices
(or with very small scattering in the color indices)
are chosen. This method calculates the ratio of
the two color excesses which can be expressed as following
\begin{equation}\label{slope}
k_{x}\equiv\frac{E(\lambda_{r}-\lambda_{x})}{E(\lambda_{c}-\lambda_{r})}
= \frac{(\lambda_{r}-\lambda_{x})_{\rm
observed}-(\lambda_{r}-\lambda_{x})_{\rm
intrinsic}}{(\lambda_{c}-\lambda_{r})_{\rm
observed}-(\lambda_{c}-\lambda_{r})_{\rm intrinsic}}
=\frac{A_{r}-A_{x}}{A_{c}-A_{r}}
\end{equation}
where $\lambda_{x}$ is the magnitude in the interested band $x$;
$\lambda_{r}$ is the magnitude in the reference band $r$
(taken to be the $\Ks$ band in this work); $\lambda_{c}$ is magnitude
in the comparison band $c$ (taken to be the J band in this work).
Therefore, the extinction ratio of the $\lambda$ band to the $\Ks$
reference band is
\begin{equation}\label{eq_ext}
A_{\lambda}/A_{\rm Ks} = 1 +
k_{\lambda}\left(1 - A_{\rm J}/A_{\rm Ks}\right) ~~,
\end{equation}
where $k_{\lambda}$ is used to derive
$A_{\lambda}/A_{\rm Ks}$ from $A_\J/A_\Ks$
and can be obtained by fitting the observed
color indices $(\Ks-\lambda)_{\rm obs}$
and $(\rm J-\Ks)_{\rm obs}$ with a linear line
\citep{Jiang03,Jiang06,Gao09}.
The slope of the fitted linear line is $k_\lambda$.
This is a statistical method as it makes use of
a large number of sources and reduces the risk of depending
on any individual objects with large uncertainties
in the determination of their intrinsic color indices.
In Figure~\ref{fig:fit}, we show the samples of fitting
for the 30 Doradus region (CO-186).\footnote{%
   Figure~\ref{fig:fit} plots $\J-\Ks$ vs. $\Ks-\lambda$.
   The slope of the fitted linear line is  $1/k_{\lambda}$.
   }

From eq.\,\ref{slope} and eq.\,\ref{eq_ext},
it is seen that the determination of $A_\lambda/A_\Ks$
(i.e., the ratio of the $\lambda$-band extinction
to the extinction of the reference band $\Ks$)
requires the knowledge of $A_{\rm J}/A_{\Ks}$.
\citet{Gordon03} obtained $A_\J/A_\Ks\approx2.96$
for the LMC\,2 supershell near 30 Doradus.
We adopt $A_\J/A_\Ks = 2.96$ since the regions
selected for this study are near 30 Doradus (CO-186)
(see Figure~\ref{fig:fields}).
For the sake of comparison, we also take the Galatcic value
of $A_\J/A_\Ks=2.52$ \citep{Rieke85} which is often
adopted to study the LMC extinction \citep{Cioni00,Imara07}.

\subsection{Tracers} \label{tracer}
In our previous work of probing the variation of
the MIR extinction law in the Galactic plane \citep{Gao09},
red giant branch stars (RGBs) were used as the tracer
to study the extinction law at the four IRAC bands.
In the IR, red giants are appropriate tracers of interstellar
extinction for the following reasons:
(i) they have a narrow range of effective temperatures
      so that the scatter of their intrinsic color indices
      is small;
(ii) they are bright in the IR ($M_{\Ks}\approx-5.0\magni$)
      and remain visible even suffering large extinction
      and/or observed from a great distance, even in the LMC.

\begin{figure}
\centering
\includegraphics[angle=0,width=3.6in]{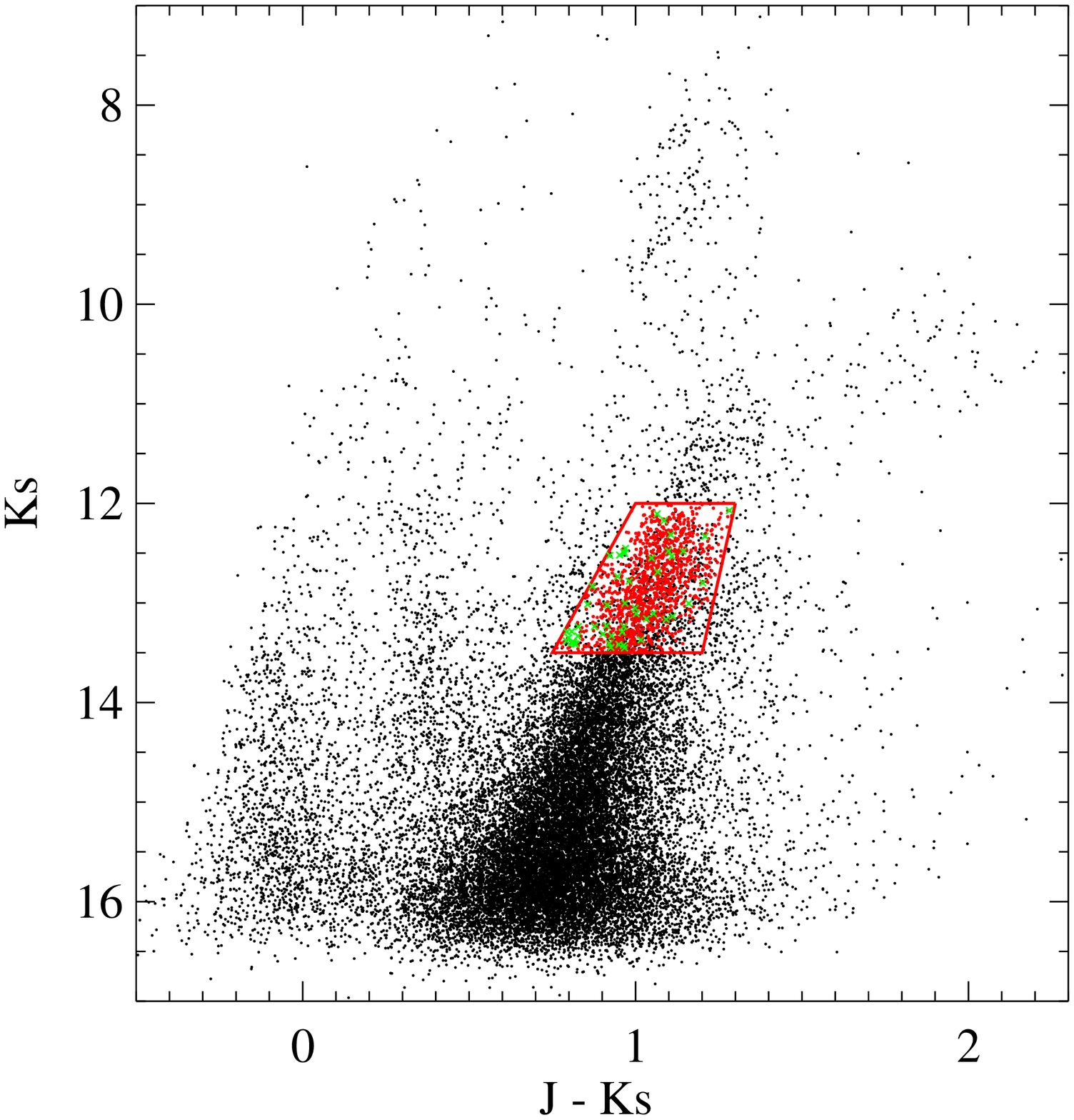}
\hspace{-0.83in}
\includegraphics[angle=0,width=3.6in]{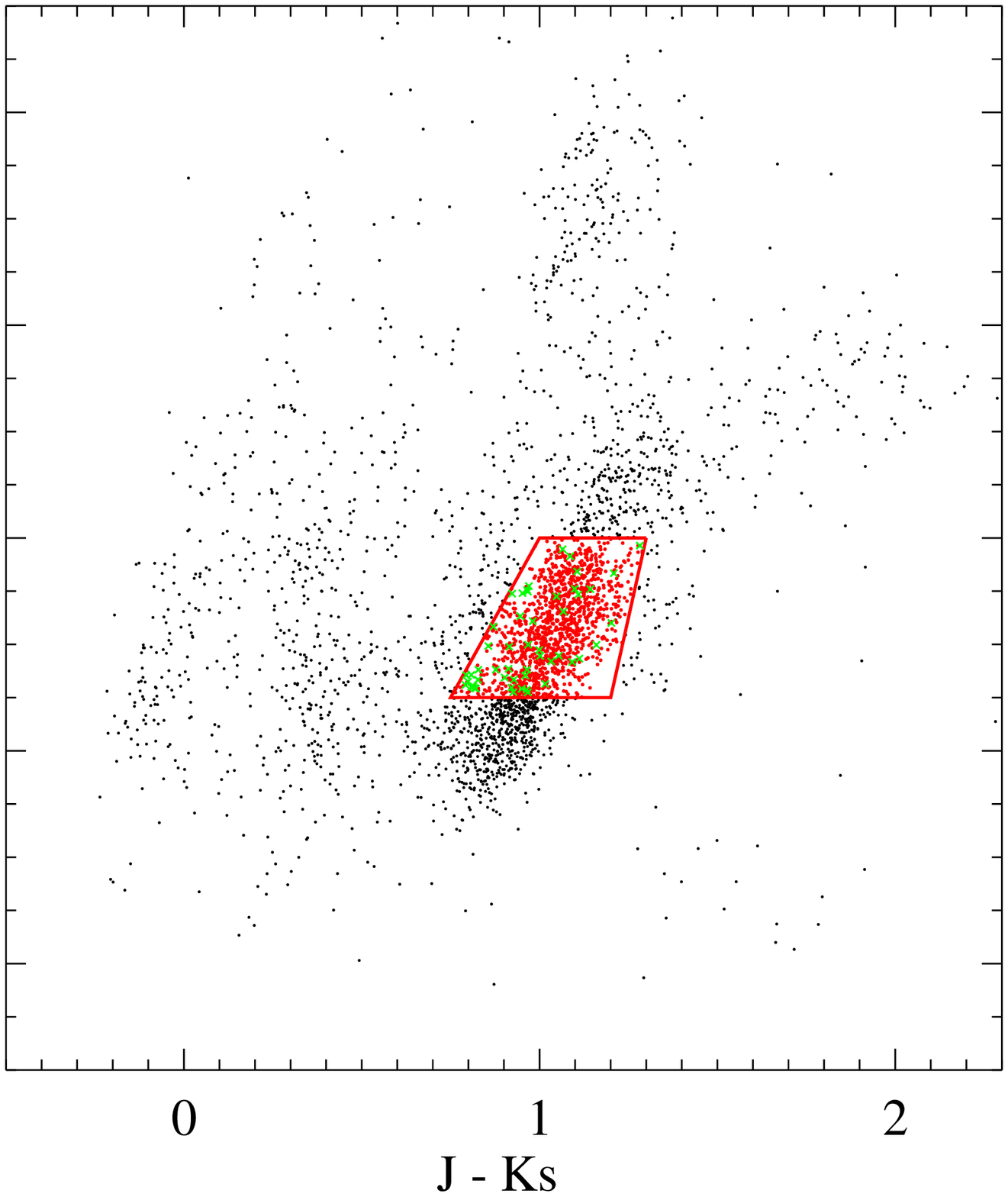}
\caption{\footnotesize
         The NIR CMD of the 30 Doradus (CO-186) region.
         The selected RGB samples are denoted by red dots
         within the red trapezoid.
         Left panel shows all the sources with S/N\,$\gtsim$\,1
         at all three 2MASS bands in the 30 Doradus (CO-186) region
         from the \emph{Spitzer}/SAGE IRAC catalog,
         while the right panel only shows the sources
         with S/N\,$\simgt$\,5 at all seven bands
         (three 2MASS bands and four IRAC bands) in the same region.
         Green crosses show the sources with $[3.6] - [4.5] > 0.0$
         or $[3.6] - [8.0] > 0.5$ in the selected region.
\label{fig:RGB186}}
\end{figure}

In the LMC, RGB stars are one of the most prominent
and well populated features
in the color-magnitude diagram (CMD) of stellar populations
with ages larger than $\simali$1.5--2.0\,Gyr \citep{Salaris05}.
\citet{Nikolaev00} performed a morphological analysis
on the 2MASS CMD of the LMC and distinguished different
populations of stars in the LMC.
The populations were identified based on isochrone fitting
and matching the theoretical CMD colors of known populations
to the observed CMD source density
(see Figure~3 and Tables~2, 3 of \citealt{Nikolaev00}).
\citet{Imara07} selected the sources
in \emph{Regions} E, F, G, H, J, L,
and part of \emph{Region} D of \citet{Nikolaev00}
to derive the extinction map of the LMC.
They eliminated \emph{Regions} A, B, C, and I,
which contain much foreground contamination.
Although it is free of foreground contamination,
\emph{Region} K is also eliminated because it consists of
dusty AGB stars whose large $\J - \K$ colors are
due to circumstellar dust.
However, the sources considered by \citet{Imara07}
contain too many different stellar populations.
We note that \emph{Region} E of \citet{Nikolaev00},
located within $12 < \Ks <13.5$ and
$0.9\simlt \rm J-Ks \simlt 1.2$,
covers the upper RGB and includes the tip of the RGB.
Therefore, we choose a slightly expanded version of
\emph{Region} E to select the RGB stars:
the selected region is constrained to
$12 \leq \Ks \leq 13.5$
and $0.75 \leq \rm \J-\Ks \leq 1.3$
(see Figures~\ref{fig:cmd},\ref{fig:RGB186}).

However, it should be noted that evolved red giants
may have circumstellar dust shells which would cause
circumstellar extinction and produce IR emission,
hence affecting our understanding of their intrinsic color indices.
Additionally, the selected region (\emph{Region} E) may contain
other populations which will affect the extinction determination.
For the MW, astronomers commonly use $[3.6]-[4.5]$\,$<$\,0.6
and $[5.8]-[8.0]$\,$<$\,0.2 as the criteria to exclude the sources
with IR excess such as pre-main-sequence stars and asymptotic
giant branch (AGB) stars \citep{Flaherty07,Gao09}.
For the LMC, \citet{Meixner06} presented initial results on the epoch~1 data
of the \emph{Spitzer}/SAGE program for a region near N79 and N83,
which is near the southwest end of the LMC bar.
They adopt a simplified point-source classification
to identify three candidate groups -- stars without dust,
dusty evolved stars, and young stellar objects (YSOs)
on the MIR color-color diagrams.
In their Figure~11 and Figure~12,
\citet{Meixner06} showed the stars
without dust are almost constrained to
$[3.6] - [4.5] < 0.0$,
$[5.8] - [8.0] < 0.5$.
Therefore, we will also use $[3.6] - [4.5] < 0.0$
and $[5.8] - [8.0] < 0.5$ in order to reduce
the contamination caused by YSOs and AGBs.\footnote{%
Distant galaxies are too faint and red
           to contaminate the RGB samples.
           We note that 90\% of the 2MASS galaxies,
           and particularly those in \emph{Region} L,
           have colors redder than $\J-\Ks =1\magni$
           \citep{Nikolaev00}.
           The typical magnitudes and colors of
           background galaxies are
           $\Ks$\,$\approx$\,16.5--19$\magni$
           and $\J-\Ks > 1.5\magni$ \citep{Kerber09,Tatton13}.
           }

In Figures~\ref{fig:fit} and \ref{fig:RGB186},
green crosses show the sources
with $[3.6] - [4.5] > 0.0$
or $[5.8] - [8.0] > 0.5$ in the selected \emph{Region} E.
Figure~\ref{fig:RGB186}
shows the NIR CMD of the 30 Doradus (CO-186) region,
which is one of the selected parts in the LMC.
The selected RGB sources are denoted by red dots
within the red trapezoid.
Left panel shows all the sources with S/N\,$\simgt$\,1
at all three 2MASS bands in the 30 Doradus (CO-186) region
from the \emph{Spitzer}/SAGE IRAC catalog,
while the right panel only shows the sources
with S/N\,$\simgt$\,5 at all seven bands
(i.e., three 2MASS bands and four IRAC bands)
in the same region.

The red clump giants (RCGs) are also often used as tracers
to probe the IR extinction in the MW.
However, RCGs in the LMC are often too faint to be observed
(at least in large number).
The tip-RGBs in the LMC have $\Ks \approx 12 \magni$ \citep{Sakai00, Nikolaev00, Salaris05, Mucciarelli06}
and RGBs are within $12 < \Ks < 16\magni$,
while RCG stars have $\Ks \approx 17 \magni$
\citep{Alves02,Mucciarelli06}.
Therefore, in this work only RGBs are considered as tracers
to probe the MIR extinction of the LMC.

\section{Results and Discussion}
\subsection{The NIR Extinction Law}
\begin{figure}
\centering
\includegraphics[angle=0,width=3.3in]{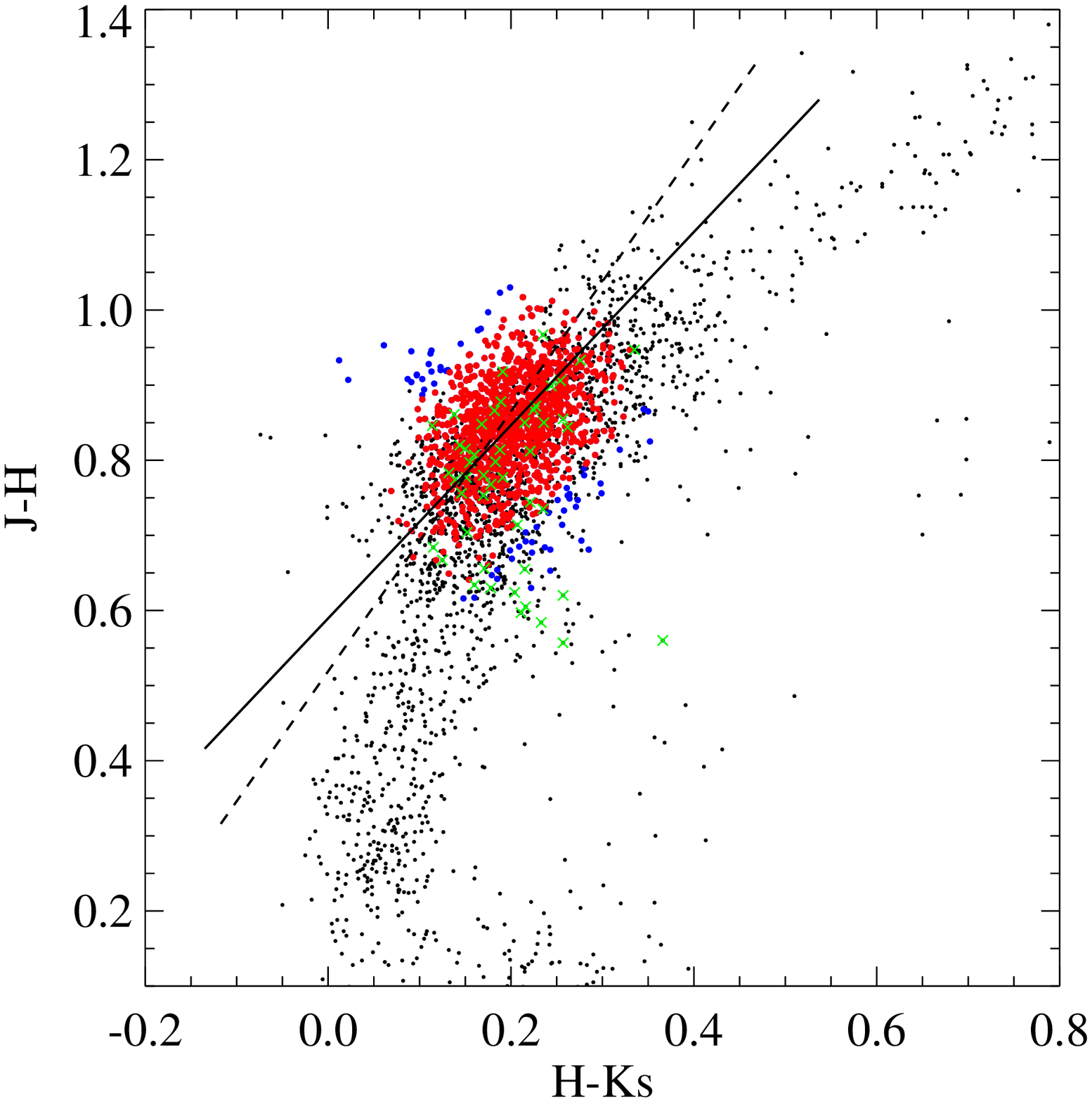}
\hspace{-0.23in}
\includegraphics[angle=0,width=3.3in]{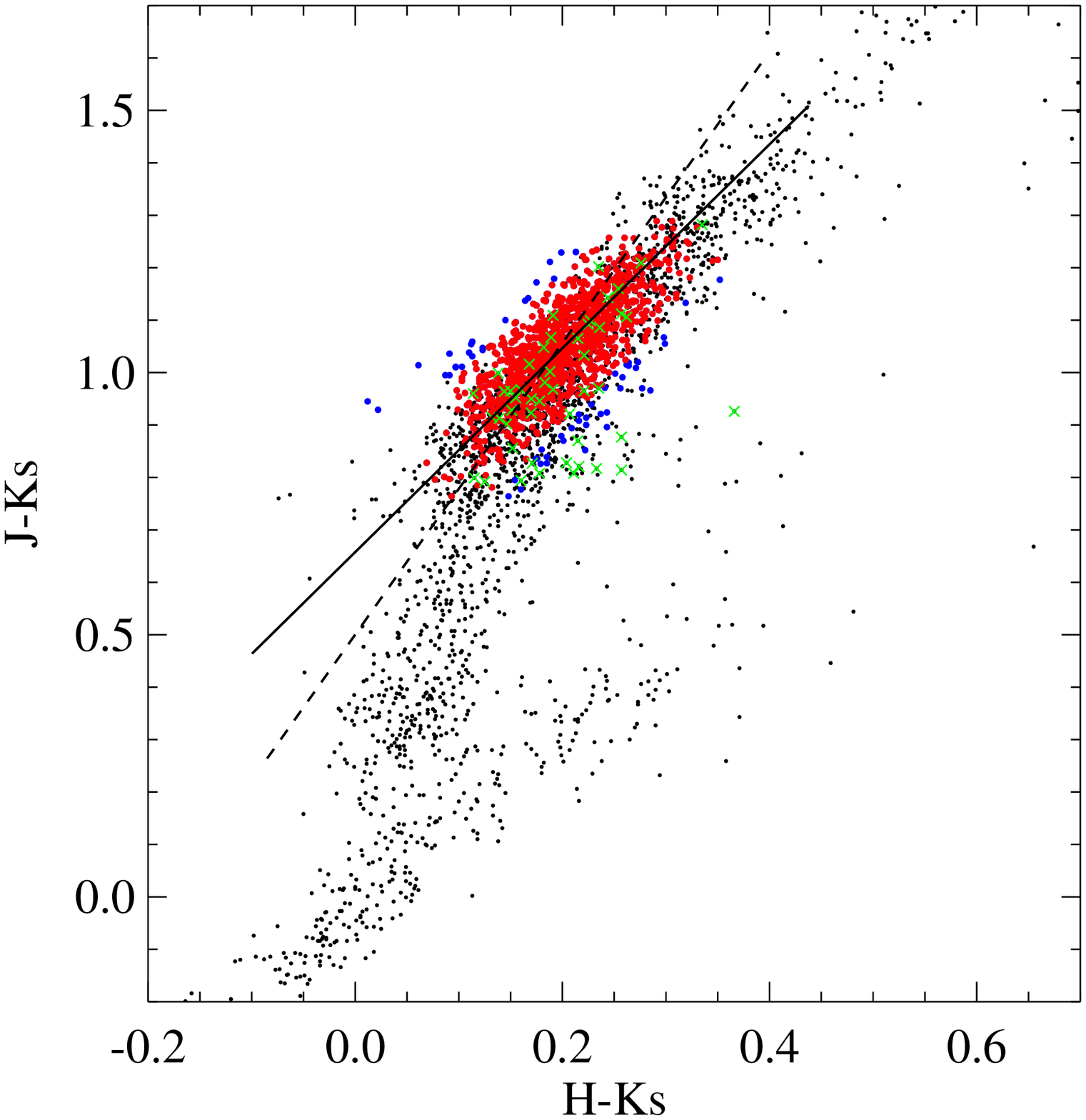}
\caption{\footnotesize
             The NIR color-color diagram of the 30 Doradus (CO-186) region.
              The symbols are the same as those
              in Figure~\ref{fig:fit}.
              For this region, the NIR color ratios are
             $E(\J-\HH)/E(\HH-\Ks) \approx 1.29\pm0.04$
             and $E(\J-\Ks)/E(\HH-\Ks) \approx 1.94\pm0.04$,
             which are shown as black solid lines.
             In the MW, these values are
            $E(\J-\HH)/E(\HH-\Ks) \approx1.73$
            and $E(\J-\Ks)/E(\HH-\Ks) \approx 2.78$
           \citep{Indebetouw05}, which are shown as black dashed lines.
\label{fig:NIR186}}
\end{figure}

\citet{Imara07} derived the NIR extinction coefficient
of the LMC using the NICE method \citep{Lada94}
and the NICER (NICE revised) method \citep{Lombardi01}.
The NICER technique relies on the mean color of
the control group being characteristic of
the intrinsic color of the field stars \citep{Imara07}.
They took the mean color of the control group
as the intrinsic color of their interested fields,
i.e., $(\HH-\K)_{\rm intrin}\equiv0.16\pm0.09\magni$.
However, their control groups include the sources in
\emph{Regions} E, F, G, H, J, L, and part of D of \citet{Nikolaev00},
so that the dispersion around the mean control color
was larger than that of RGB stars.
Based on the selected RGB stars,
we fit the NIR color-color diagrams of
$\J-\HH$ versus $\HH-\Ks$
and $\J-\Ks$ versus $\HH-\Ks$
with the IDL robust linear fit procedure.
Our results show that the average NIR extinction coefficient
for the first five LMC regions listed in Table~\ref{tab:field}
is consistent with previous studies,
i.e., $E(\J-\HH)/E(\HH-\Ks) \approx 1.29\pm0.04$
and $E(\J-\Ks)/E(\HH-\Ks) \approx 1.94\pm0.04$
(see Table~\ref{tab:IR_ext}).
In Figure~\ref{fig:NIR186},
we show the NIR color-color diagrams
and the fits for the 30 Doradus region (CO-186).
Green crosses show the sources with $[3.6] - [4.5] > 0.0$
or $[3.6] - [8.0] > 0.5$ in the selected \emph{Region}
E of 30 Doradus. It efficiently reduces the contamination
of YSOs and AGBs, and prevents from distorting
the fit by excluding these sources.

As far as individual parts of the LMC are concerned,
the NIR extinction coefficients
vary among different regions of the LMC.
However, the variation is relatively small among
the first five regions listed in Table~\ref{tab:field}
which have molecular clouds and for which $\AV$ is large.
For comparison, Table~\ref{tab:IR_ext} also shows
the NIR extinction coefficients
for the LMC bar and HI region.
Because of their small amounts of extinction,
we believe that the results (for the LMC bar and HI region)
are less certain (compared to those
for the first five regions listed in Table~\ref{tab:field})
and the IR extinction cannot
be detected efficiently.

\subsection{The MIR Extinction Law\label{sec:mir}}
In Table~\ref{tab:IR_ext}, the color ratios of
$E(\Ks-\lambda)/E(\J-\Ks)$ are shown
for the four IRAC bands.
Based on these color ratios
and taking $A_{\J}/A_{\Ks} = 2.96$ \citep{Gordon03},
we calculate the relative extinction $A_{\lambda}/A_{\Ks}$.
In Table~\ref{tab:MIR_ext} we show the extinction
at each IRAC band (relative to that of the $\Ks$ band)
for the selected regions of the LMC.
We note that only the first five regions listed in Table~1
are considered in obtaining the average MIR extinction of the LMC
(although the MIR extinction toward the LMC bar
and HI region are also derived and tabulated
in Table~\ref{tab:MIR_ext}).

\begin{figure}
\epsscale{.70} \plotone{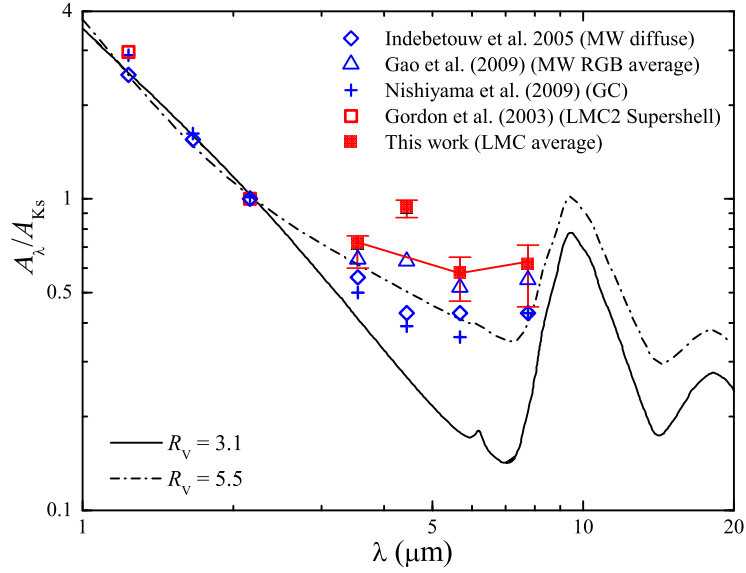}
\centering
\caption{\footnotesize
          The IR extinction law of the LMC
          compared with the MW model extinction curves
           \citep{Weingartner01}
          of $\Rv = 3.1$ (black solid line)
          and $\Rv = 5.5$ (black dot-dashed line).
          Red open squares:
                  the LMC NIR extinction of \citet{Gordon03}.
          Red filled squares: the MIR extinction at the four IRAC bands
                              derived here, and the error bars
                              show the maximum and minimum extinction
                              among the five selected fields.
          Blue open diamonds: the NIR and MIR extinction
                              for the MW diffuse ISM
                              of \citet{Indebetouw05}.
          Blue filled triangles: the average extinction
                                 at the four IRAC bands
                                 derived from 131 GLIMPSE fields
                                 along the Galactic plane
                                 with RGBs as tracers \citep{Gao09}.
          Blue pluses: the NIR and MIR extinction
                       toward the Galactic Center
                       of \citet{Nishiyama09}.
                       With the possible systematic bias
                       (caused by the correlated uncertainties of $\Ks-\lambda$ and $\J-\Ks$) corrected,
                       the MIR extinction $A_\lambda/A_\Ks$ decreases slightly at all bands, but it remains flat and excessive at [4.5]$\mum$.
\label{fig:ext_curve}}
\end{figure}

As a whole, with RGBs as the extinction tracers,
the mean MIR extinction ratios of the LMC are
$A_{[3.6]}/A_\Ks\approx0.72\pm0.03$,
$A_{[4.5]}/A_\Ks\approx0.94\pm0.03$,
$A_{[5.8]}/A_\Ks\approx0.58\pm0.04$, and
$A_{[8.0]}/A_\Ks\approx0.62\pm0.05$.
In Figure~\ref{fig:ext_curve}, we show the MIR extinction
of the LMC together with the MW model extinction curves
of $\Rv=3.1$ and $\Rv=5.5$ \citep{Weingartner01}.
The red filled squares show the average MIR extinction
of the LMC at the four IRAC bands,
while the red error bars show the maximum and minimum extinction
of the five selected fields, demonstrating the regional variation
of MIR extinction in the LMC.
If the extinction at [4.5]$\mum$ is excluded (see \S4.3),
the extinction at [3.6], [5.8] and [8.0]$\mum$ of the LMC
consists a flat curve, close to that of the MW model extinction
curve of $\Rv = 5.5$,
lacking the deep minimum around 7$\mum$ predicted from
the $\Rv=3.1$ model.
The extinction at [4.5]$\mum$ is well above
the $\Rv=5.5$ model extinction curve and
also well exceeds the extinction at the other three IRAC bands.
As far as individual sightlines are concerned,
the wavelength dependence of the LMC MIR extinction
$A_{\lambda}/A_{\Ks}$ also varies from one sightline
to another (see Table~\ref{tab:IR_ext} and \ref{tab:MIR_ext}).

In the literature the NIR extinction law of the MW has
often been used to represent that of the LMC
\citep{Cioni00,Imara07}.
For comparison, we have also calculated the MIR extinction
of the LMC taking the Galactic value of
$A_\J/A_\Ks=2.52$ \citep{Rieke85}.
The results are also tabulated in Table~\ref{tab:MIR_ext}:
the extinction is larger
than that derived with the LMC2 supershell value
of $A_\J/A_\Ks=2.96$ \citep{Gordon03}
by $\simali$8\%, 2\%, 17\%, and 15\%
at [3.6], [4.5], [5.8], and [8.0], respectively.

\subsection{The Extinction at [4.5]$\mum$\label{ice45}}
As clearly seen in Figure~\ref{fig:ext_curve}
and Table~\ref{tab:MIR_ext},
the mean extinction of the 4.5$\mum$ IRAC band
$A_{[4.5]}/A_{\Ks}\approx0.94\pm0.03$
is not only much higher than that of the other three IRAC bands,
but also much larger than that of the Galactic plane average
of $A_{[4.5]}/A_{\Ks}\approx0.63\pm0.01$
derived from the RGBs tracers \citep{Gao09}.

In Galactic and extragalactic dense clouds
various ice features have been observed,
including the 3.1$\mum$ feature due to the O--H stretching
mode of H$_2$O ice as well as a number of weaker features
at 4.27$\mum$ due to the C--O stretching mode of CO$_2$ ice,
at 4.67$\mum$ due to the C--O stretching mode of CO ice,
and at 6.02$\mum$ due to the H--O--H bending mode of H$_2$O ice.
In this work, the selected sightlines are mostly
located in molecular clouds, therefore,
it is not unreasonable, at a first glance,
to attribute the [4.5] excess extinction to
$\rm CO_2$ and CO ices in the LMC dense clouds.
Observational, experimental and theoretical
studies have shown that CO and CO$_2$ ices could
efficiently form both in quiescent clouds and
in active star-forming regions
(e.g., see \citealt{Nummelin01,Ruffle01,Mennella04,
Whittet07,Oba10,Noble11,Ioppolo11,Garrod11}).
\citet{Whittet07} found a threshold of
$\Av$\,$\approx$\,4.3$\pm$1.0, 6.7$\pm$1.6$\magni$
respectively for the detection of CO$_2$ and CO ices
in the Taurus dark cloud complex;
for H$_2$O ice, the detection threshold is
$\Av$\,$\approx$\,3.2$\pm$0.1$\magni$
\citep{Whittet01b}. With $\Av<5\magni$ for the selected LMC regions,
it is unlikely for these regions to have CO$_2$/H$_2$O or CO/H$_2$O
much higher than that of the MW.

To examine whether the ice (particularly CO$_2$
and, to a less degree, CO) absorption bands could account for
the [4.5] excess extinction, we approximate the 3.05, 4.27, 4.67
and 6.02$\mum$ bands respectively from H$_2$O, CO$_2$, CO,
and H$_2$O ices as four Drude profiles.\footnote{%
  We do not consider the interstellar 4.62$\mum$
       ``XCN'' absorption feature. This feature is commonly
       seen toward luminous protostars embedded
       in dense molecular clouds.
       Its carrier lacks a specific identification,
       although it is generally believed that the
       4.62$\mum$ feature arises from the C$\equiv$N stretch
       of some sort of CN-bearing organic dust
       and the carrier may result from energetic processing
       (e.g., UV photolysis or ion bombardment) of interstellar
       ice mixtures containing N in the form of NH$_3$ or N$_2$
       \citep{Lacy84,Schutte97,Pendleton99,Whittet01a}.
        The ``XCN'' abundance is small, with
        XCN/H$_2$O\,$<$\,6\% in high-mass young stellar objects
        which have the highest XCN abundance (see \citealt{Whittet01a}).
        }
The extinction due to these four ice bands is
\begin{equation}\label{eq:ice}
A_{\lambda}({\rm ice}) = 1.086\,N_{\rm H_{2}O} \sum\limits_{j=1}^4
\frac{A_j \times \left[{\rm X}_j/{\rm H_{2}O}\right] \times E_j
\times 2\gamma_j/\pi}
{\left(\lambda -\lambda_j^2/\lambda\right)^2+\gamma_j^2}
\end{equation}
where $\lambda_j$, $\gamma_j$, $A_j$ are the peak wavelength,
FWHM, and strength of the $j$-th ice absorption band, respectively
(see Table~\ref{tab:ice});\footnote{%
   The FWHM $\gamma_j$ and band strength $A_j$
   values tabulated in Table~\ref{tab:ice} are in units of
   cm$^{-1}$ and cm\,molecule$^{-1}$, respectively.
   When using eq.\,\ref{eq:ice},
   they are converted so that they are in units of
   $\mum$ and  cm$^3$\,molecule$^{-1}$.
    }
$N_{\rm H_2O}$ is the H$_2$O ice column density;
${\rm X}_j/{\rm H_2O}$ is the abundance of
the ice species ${\rm X}_j$ relative to H$_2$O ice
in typical dense clouds;
$E_j$ is the ``enhancement'' factor for species ${\rm X}_j$
(i.e., ${\rm X}_j/{\rm H_2O}$ is increased by
a factor of $E_j$ in order for the ice absorption bands
to account for the [4.5] excess extinction).
Finally, we add $A_\lambda({\rm ice})$ to
$\left(A_\lambda/A_{\Ks}\right)_{\rm WD}$,
the WD01 extinction curve (with $\Rv=3.1$ or 5.5) \citep{Weingartner01}
and then convolve with the {\it Spitzer}/IRAC
filter response functions.\footnote{%
   Since the ice bands contribute little to the extinction at the
   $\Ks$ band, the addition of the ice extinction does not change
   the normalization of $A_{\Ks}$.
   But in any case, we re-normalize the final extinction at $\Ks$.
   }

\begin{figure}
\centering
\includegraphics[angle=0,width=4.8in]{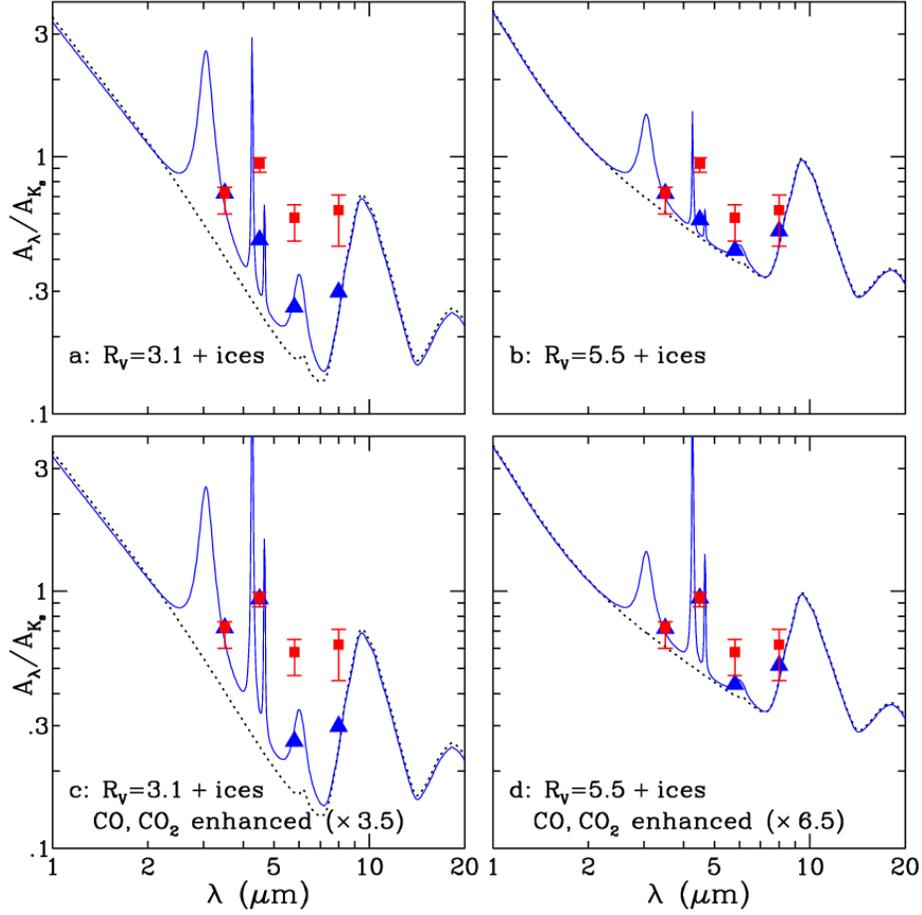}
\caption{\footnotesize
         Comparison of the LMC mean MIR extinction
         at the four IRAC bands (red filled squares)
         with the $\Rv=3.1$ and $\Rv=5.5$ models
         (black dotted lines)
         combined with the 3.05 and 6.02$\mum$ absorption
         bands of H$_2$O ice,
         the 4.27$\mum$ band of CO$_2$ ice,
         and the 4.67$\mum$ band of CO ice
         (blue solid lines).
         Blue filled triangles are
         the $\Rv=3.1$ or $\Rv=5.5$ model curve
         plus the ice absorption bands convolved
         with the {\it Spitzer}/IRAC filter functions.
         Upper panels (a,\,b): the abundances of
         CO and CO$_2$ ices are taken to be that
         of typical dense clouds
         (i.e., ${\rm CO/H_2O =0.25}$,
          ${\rm CO_2/H_2O =0.21}$).
         Bottom panels (c,\,d): the abundances of
         CO and CO$_2$ ices are enhanced relative
         to their typical values in dense clouds.
         }
\label{fig:ice}
\end{figure}

The H$_2$O ice abundance (i.e., $N_{\rm H_2O}$)
is constrained not to exceed the IRAC [3.6] extinction.
With ${\rm CO/H_2O \approx 0.25}$
and ${\rm CO_2/H_2O \approx 0.21}$,
typical for quiescent dense clouds \citep{Whittet03,Gibb04,Boogert11},
CO and CO$_2$ are not capable of accounting for
the [4.5] excess extinction.
If we are forced to attribute the [4.5] excess extinction
to CO and CO$_2$, we will have too much H$_2$O ice
and the resulting extinction at [3.6] would be too high.
If we fix the H$_2$O ice abundance at what is required by
the [3.6] extinction, in order for CO and CO$_2$ to explain
the [4.5] excess extinction, we have to enhance their abundances
(relative to their typical abundances in dense clouds)
by a factor of
$E_{\rm CO}=E_{\rm CO_2}\approx 3.5$ for $\Rv=3.1$
and $E_{\rm CO}=E_{\rm CO_2}\approx 6.5$ for $\Rv=5.5$
(see Figure~\ref{fig:ice}).

We note that although the extinction at [3.6] and [4.5]
is well fitted by the $\Rv=3.1$ model combined with ices,
it could not account for the flat extinction at [5.8] and [8.0].
In contrast, the $\Rv=5.5$ model together with ices
could closely explain the MIR extinction at all four
IRAC bands provided that the abundances of CO and CO$_2$ ices
are increased by a factor of $\simali$6.5
from their typical abundances in dense clouds.
However, the required CO and CO$_2$ abundances are
unrealistically too high:
${\rm CO/H_2O \approx 1.7}$
and ${\rm CO_2/H_2O \approx 1.4}$
for $\Rv=5.5$.\footnote{%
\citet{Shimonishi08} reported the detection of
   H$_2$O and CO$_{2}$ ices toward massive YSOs
   in the LMC based on the AKARI LMC spectroscopic survey. They estimated ${\rm CO/H_2O \leq 0.19}$
   and ${\rm CO_2/H_2O \approx 0.36\pm0.09}$
   (see their Table 4).
   }

The origin of the [4.5] excess extinction remains unclear.
Some red giants have circumstellar envelopes rich in CO gas
which absorbs at 4.6$\mum$ \citep{Bernat81}.
With red giants as a tracer of the mid-IR extinction,
the 4.6$\mum$ absorption feature of their CO gas
would result in an overestimation of the IRAC [4.5] extinction.
It is worth exploring whether the CO gas absorption of red giants
could account for the [4.5] excess extinction.

\subsection{Systematic Bias}
As discussed in \S\ref{sec:cem},
in deriving the IR extinction law,
the $\Ks-\lambda$ vs. $\J-\Ks$ color-color diagram
is fitted to obtain the slope $k_\lambda$
(see eq.\ref{slope} and eq.\ref{eq_ext}).
Since $\Ks-\lambda$ and $\J-\Ks$ both employ
the $\Ks$ band photometry and therefore may have
correlated uncertainties, there could be systematic bias
in determining $k_\lambda$ and $A_\lambda/A_\J$.

\begin{figure}
\centering
\includegraphics[angle=0,width=5in]{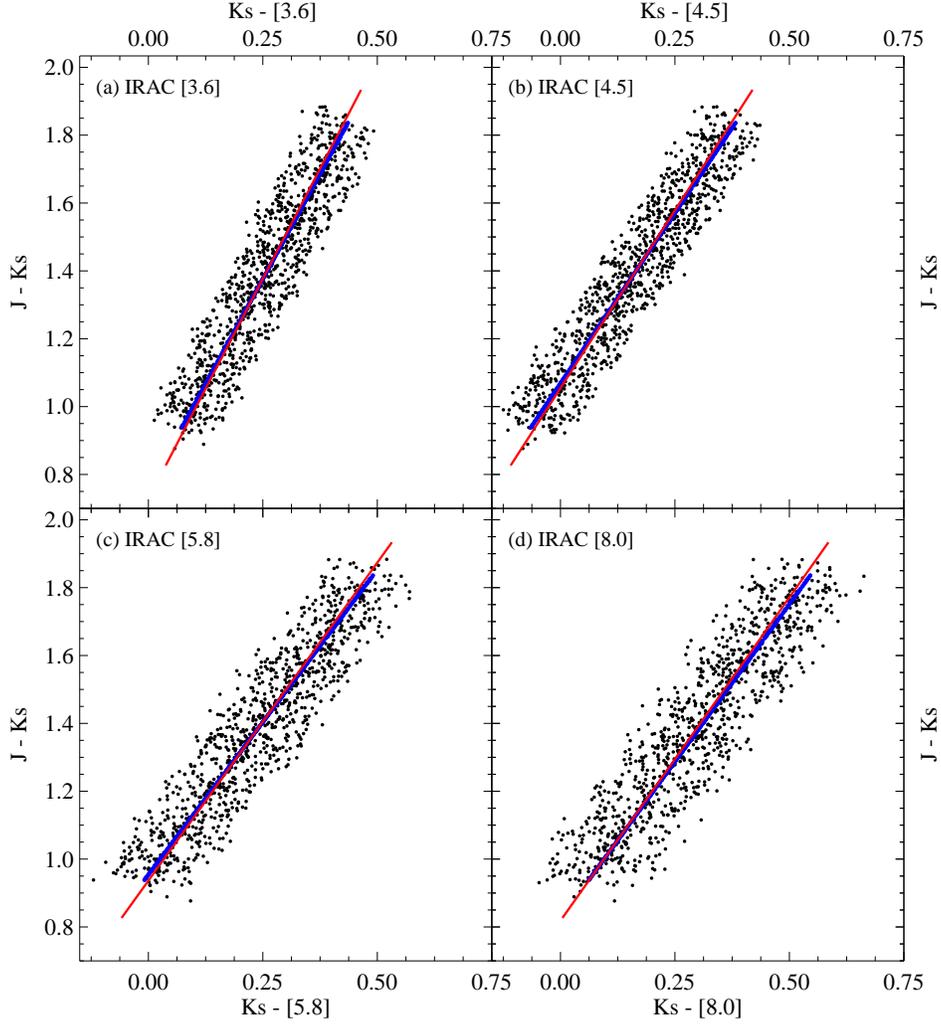}
\caption{\footnotesize
         A sample of Monte-Carlo simulation
         with $N=1000$, $\Rv=3.1$,
              $\left(A^{\rm mc}_{\J}\right)_{\rm max}=1.5\magni$,
         and the mean photometric uncertainties
         for the sources with $\StoN>5$ in all
         three 2MASS bands and four IRAC bands.
         The thick blue lines show the simulated colors
         of the obscured sources
         (but no errors are added to these sources).
         The black dots show the colors of
         the simulated, obscured, error-added sources.
         The red lines fit the $\J-\Ks$ vs. $\Ks-\lambda$
         color-color diagram for
         the simulated, obscured, error-added sources.
         \label{fig:montecarlo}}
\end{figure}

To evaluate the possible systematic bias,
we perform simple Monte-Carlo simulations.
Using CMD 2.5 \citep{Bressan12}\footnote{%
  {\sf http://stev.oapd.inaf.it/cgi-bin/cmd}
  },
we generate stellar isochrones and take a stellar
atmosphere model of
$T_{\rm eff}=4,004\K$, $Z=0.019$, ${\rm log\,g}\,=\,1.28$
for the simulated stars.
The stellar absolute magnitudes are approximatly
$-4.068$, $-4.850$, $-5.006$, $-5.079$,
$-4.941$, $-4.997$, and $-5.068\magni$
in $\J$, $\HH$, $\Ks$, [3.6], [4.5], [5.8],
and [8.0] bands, respectively.
We then apply a random amount of extinction
to obscure a large number of stars.
The stars are taken to be identical and have
the same absolute magnitudes.
Let $A^{\rm mc}_\lambda$ be
the amount of extinction to which the stars are subject.
We take the stars to be obscured by $A^{\rm mc}_\lambda$,
with $A^{\rm mc}_\lambda$ being a random number between
0 and $\left(A^{\rm mc}_\lambda\right)_{\rm max}$.
We take the $\J$ band as the basis wavelength
and assume the extinction at other wavelengths
to follow either the MW $\Rv=3.1$ model curve or
the MW $\Rv=5.5$ model curve of \citet{Weingartner01}.
The thick blue lines in Figure~\ref{fig:montecarlo}
shows the simulated color indices
for a large number of stars ($N=1,000$)
in the $\J-\Ks$ vs. $\Ks-\lambda$ color-color diagram.
The stars are obscured by the $\Rv=3.1$-type extinction
with $\left(A^{\rm mc}_\J\right)_{\rm max}=1.5\magni$.

\begin{figure}
\centering
\includegraphics[angle=0,width=7in]{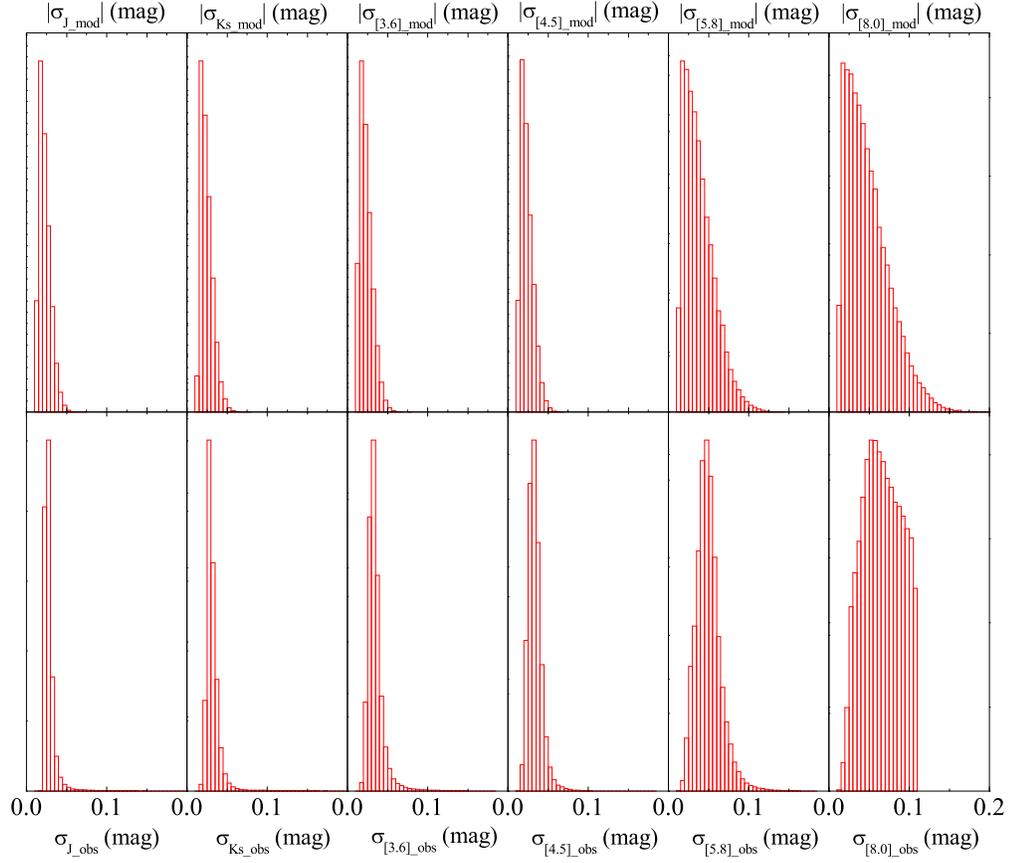}
\caption{\footnotesize
         Comparison of the simulation errors (upper panel)
         with the photometric uncertainties
         for the sources with $\StoN>5$ at all
         three 2MASS bands and four IRAC bands
         (lower panel; see Table~\ref{tab:obserr}).
\label{fig:obserr}}
\end{figure}

We then add errors to the obscured stellar magnitudes
based on the uncertainties of the SAGELMCcatalogIRAC
(see Table~\ref{tab:obserr}).
The errors are simulated to follow a Poisson
distribution, with their mean values being
the mean photometric uncertainties of
the sources with $\StoN>5$
in all seven bands in SAGELMCcatalogIRAC
(see Figure~\ref{fig:obserr} for a comparison of
the simulated errors with the mean photometric
uncertainties).
In Figure~\ref{fig:montecarlo},
the simulated, error-added sources are shown as black dots
in the $\J-\Ks$ vs. $\Ks-\lambda$ color-color diagram.
The slope for the resulting $\J-\Ks$ vs. $\Ks-\lambda$
diagram (black dots) is fitted
(see the red lines in Figure~\ref{fig:montecarlo}).
Figure~\ref{fig:montecarlo}
shows that the extinction laws derived from
the obscured, error-added sources are different from
the original ones adopted to obscure the stars:
the derived slopes of $k_{\lambda}$,
 i.e. $E(\Ks-\lambda)/E(\J-\Ks)$,
are all smaller than the simulated ones,
and therefore the derived $A_{\lambda}/A_{\rm Ks}$ ratios
are overestimated.

We have examined the effects of
(i) the level of obscuration
$\left(A^{\rm mc}_\J\right)_{\rm max}$,
(ii) the number of stars $N$,
(iii) the amount of error, and
(iv) the extinction-type which is employed
to obscure the stars.
As shown in Table~\ref{tab:montecarlo},
the level of obscuration
$\left(A^{\rm mc}_\J\right)_{\rm max}$
dominates the systematic bias
which decreases with the increasing
of $\left(A^{\rm mc}_\J\right)_{\rm max}$.
For $\left(A^{\rm mc}_\J\right)_{\rm max}\simgt5\magni$
the systematic bias becomes negligible.

\citet{Gordon03} derived
$A_\J/\Av\approx0.299$ for the lines of sight
toward the LMC2 supershell near 30 Doradus.
With $\Av\simlt5\magni$, we obtain
$A_\J\simlt1.5\magni$.
Therefore, the MIR $A_{\lambda}/A_{\rm Ks}$ ratios
may have been overestimated and
the color ratios $E(\Ks-\lambda)/E(\J-\Ks)$
may have been underestimated.
With $N=1,000$,
$\left(A^{\rm mc}_\J\right)_{\rm max} = 1.5\magni$,
and a $\Rv=3.1$-type extinction (see the fifth row in Table~\ref{tab:montecarlo}),
we perform a Monte-Carlo simulation to assess
the possible systematic bias caused by
the correlated uncertainties of
$\Ks-\lambda$ and $\J-\Ks$.
We find that the extinction ratios
$A_\lambda/A_\J$ derived in \S\ref{sec:mir}
could be overestimated by
$\simali$6\%, 10\%, 16\% and 9\%
at [3.6], [4.5], [5.8] and [8.0], respectively.
The corrected extinction $A_{\lambda}/A_{\rm Ks}$ will be
$A_{[3.6]}/A_\Ks\approx0.68\pm0.03$,
$A_{[4.5]}/A_\Ks\approx0.84\pm0.03$,
$A_{[5.8]}/A_\Ks\approx0.49\pm0.04$, and
$A_{[8.0]}/A_\Ks\approx0.57\pm0.05$.


\section{Conclusions}\label{conclusions}
The \emph{Spitzer}/SAGE IRAC catalog provides us
a unique opportunity to study the MIR extinction of the LMC.
We select five fields with large optical extinction ($\Av >1\magni$)
to explore the MIR extinction of the sightlines toward these regions.
Taking $A_\J/A_\Ks = 2.96$ \citep{Gordon03}
and using RGB stars as tracers, we find that:
\begin{enumerate}
\item The NIR color-color diagrams are fitted
      to derive the NIR extinction coefficients:
      $E(\J-\HH)/E(\HH-\Ks) \approx 1.29\pm0.04$
      and $E(\J-\Ks)/E(\HH-\Ks) \approx 1.94\pm0.04$,
      which are consistent with earlier studies
      \citep{Koornneef82, Imara07}.
      The corresponding color ratios
       for the MW are
       $E(\J-\HH)/E(\HH-\Ks) \approx 1.73$
       and $E(\J-\Ks)/E(\HH-\Ks) \approx 2.78$ \citep{Indebetouw05}.

\item At the four IRAC bands,
      the derived mean MIR extinction
      of the LMC are $A_{[3.6]}/A_\Ks\approx0.72\pm0.03$,
      $A_{[4.5]}/A_\Ks\approx0.94\pm0.03$,
      $A_{[5.8]}/A_\Ks\approx0.58\pm0.04$, and
      $A_{[8.0]}/A_\Ks\approx0.62\pm0.05$. The corresponding extinction ratios
       for the MW are
       $A_{[3.6]}/A_\Ks\approx0.63\pm0.01$,
       $A_{[4.5]}/A_\Ks\approx0.57\pm0.03$,
       $A_{[5.8]}/A_\Ks\approx0.49\pm0.03$, and
       $A_{[8.0]}/A_\Ks\approx0.55\pm0.03$ \citep{Gao09}.

\item The LMC extinction at [3.6], [5.8] and [8.0]$\mum$
      is consistent with a flat curve, close to that of
      the MW model extinction curve predicted by
      the interstellar grain model of $\Rv = 5.5$.
      Similar to that of the MW, the LMC MIR extinction law
      exhibits appreciable regional variations.

\item The extinction at [4.5] is much higher than
      that of the other three IRAC bands.
      It cannot be explained in terms of
      the 4.27$\mum$ absorption band of $\rm CO_2$ ice
      and the 4.67$\mum$ absorption band of CO ice.
      It may be caused by the 4.6$\mum$ absorption feature
      of CO gas in the circumstellar envelopes of
      red giants which are used to trace the IR extinction.

\item The derived MIR extinction $A_\lambda/A_\J$
      may be overestimated because of the correlated
      uncertainties of $\Ks-\lambda$ and $\J-\Ks$ which
      could affect the determination of $k_\lambda$,
      the slope of the $\Ks-\lambda$ vs. $\J-\Ks$
      color-color diagram.
      With this systematic bias taken into account,
      the derived extinction ratios $A_\lambda/A_\J$
      could be overestimated by $\simali$6\%, 10\%, 16\%
      and 9\% at [3.6], [4.5], [5.8] and [8.0], respectively.

\end{enumerate}

\acknowledgments{%
We thank the anonymous referee
for his/her very helpful suggestions.
We thank A.C.A.~Boogert and E.~Gibb for helpful
discussions. This work is supported by
NSF AST-1109039, NNX13AE63G,
NSFC\,11173007, NSFC\,11173019, NSFC\,11273022,
the University of Missouri Research Board,
and the John Templeton Foundation in conjunction
with National Astronomical Observatories,
Chinese Academy of Sciences.
}

{\it Facilities:} \facility{\emph{Spitzer}(IRAC)}, \facility{2MASS}.


\begin{table}[htbp]
\begin{center}
\caption{The mean photometric uncertainties
         in SAGELMCcatalogIRAC\tablenotemark{a}\label{tab:obserr}}
\vspace{0.2in}
\begin{tabular}{c|ccccccc}
\hline\hline
$\lambda$                 &  $\J$  & $\HH$  & $\Ks$  & [3.6]  & [4.5]  & [5.8]  & [8.0] \\  \hline
$\langle \sigma\rangle$ ($\magni$)   &0.0280  &0.0296  &0.0316  &0.0343  &0.0330  &0.0492  &0.0658 \\
\hline
\end{tabular}
\tablenotetext{a}{Sources with $\StoN>5$
                  at all three 2MASS and four IRAC bands}
\end{center}
\end{table}

\begin{table}
\begin{center}
\caption{The LMC sightlines explored in this work \label{tab:field}}
\vspace{0.2in}
\begin{tabular}{l|cccccl}
\tableline \tableline
No.   &    Field Name   &    $\rm RA_{min}$
      &  $\rm RA_{max}$   &   $\rm DEC_{min}$
      & $\rm DEC_{max}$  & Note\tablenotemark{a} \\
      &     &  (deg) &  (deg) & (deg) &  (deg) &  \\
\tableline
1     & CO-154          &      82.5            &      84.0         &       -68.7        &    -68.0        &  CO Cloud 154 \\
2     & 30 Doradus      &      83.5            &      86.0         &       -69.5        &    -68.7        &  Star-Forming Region, CO Cloud 186\\
3     & MR North        &      84.0            &      86.0         &       -70.0        &    -69.5        &  Molecular Ridge North, CO Cloud 191 \\
4     & MR South        &      84.0            &      86.0         &       -71.0        &    -70.0        &  MR Center and South, CO Cloud 197 \\
5     & Arcs            &      86.0            &      88.0         &       -70.5        &    -69.0        &  Arc II and Arc III, CO Cloud 216, 226\\
\hline
6     & LMC Bar         &      80.0            &      82.0         &       -70.0        &    -69.0        &  CO Cloud 133, 105\\
7     & HI Region       &      86.0            &      88.0         &       -68.5        &    -67.5        &  HI Region\\
\tableline
\tableline
\end{tabular}
\tablenotetext{a}{CO cloud number: taken from \citet{Fukui08}
                  and \citet{Dobashi08}}
\end{center}
\end{table}

\begin{table}  
\small
\begin{center}
\caption{The IR extinction coefficients of the LMC}\label{tab:IR_ext}
\vspace{0.2in}
\begin{tabular}{l|cc|cccc}
\hline\hline
Field  & \large{$\frac{E(\J-\HH)}{E(\HH-\Ks)}$}    &  \large{$\frac{E(\J-\Ks)}{E(\HH-\Ks)}$}
       & \large{$\frac{E(\Ks-[3.6])}{E(\J-\Ks)}$}  &  \large{$\frac{E(\Ks-[4.5])}{E(\J-\Ks)}$}
       & \large{$\frac{E(\Ks-[5.8])}{E(\J-\Ks)}$}  &  \large{$\frac{E(\Ks-[8.0])}{E(\J-\Ks)}$} \\
\hline
CO-154        & 1.310$\pm$0.053& 1.936$\pm$0.049& 0.145$\pm $0.018& 0.031$\pm $0.020& 0.211$\pm $0.025& 0.204$\pm $0.036 \\  
30 Doradus    & 1.286$\pm$0.039& 1.944$\pm$0.037& 0.124$\pm $0.013& 0.028$\pm $0.014& 0.202$\pm $0.019& 0.187$\pm $0.029 \\  
MC North      & 1.271$\pm$0.046& 1.934$\pm$0.042& 0.206$\pm $0.016& 0.068$\pm $0.012& 0.269$\pm $0.024& 0.280$\pm $0.035 \\  
MC South      & 1.279$\pm$0.026& 1.932$\pm$0.025& 0.120$\pm $0.009& 0.006$\pm $0.009& 0.180$\pm $0.012& 0.150$\pm $0.018 \\  
Arcs          & 1.310$\pm$0.030& 1.957$\pm$0.029& 0.124$\pm $0.010& 0.013$\pm $0.010& 0.196$\pm $0.013& 0.146$\pm $0.018 \\  
\hline
Average       & 1.291$\pm$0.039& 1.941$\pm$0.036& 0.144$\pm $0.013& 0.029$\pm $0.013& 0.212$\pm $0.019& 0.193$\pm $0.027\\  %
\hline
HI region     & 1.138$\pm$0.050& 1.706$\pm$0.047& 0.054$\pm $0.018&-0.078$\pm $0.019& 0.087$\pm $0.026& 0.075$\pm $0.033 \\  
LMC bar       & 1.185$\pm$0.019& 1.782$\pm$0.018& 0.125$\pm $0.007&-0.025$\pm $0.007& 0.192$\pm $0.010& 0.166$\pm $0.013 \\  
\hline\hline
\end{tabular}
\end{center}
\normalsize
\end{table}

\begin{table}  
\small
\begin{center}
\caption{The MIR extinction law of the LMC\tablenotemark{a}}\label{tab:MIR_ext}
\vspace{0.2in}
\begin{tabular}{l|cccc}
\hline\hline
Field & $A_{[3.6]}/A_\Ks$ & $A_{[4.5]}/A_\Ks$  & $A_{[5.8]}/A_\Ks$ & $A_{[8.0]}/A_\Ks$ \\
\hline
CO-154        & 0.72$\pm $0.04 & 0.94$\pm $0.04 & 0.59$\pm $0.05& 0.60$\pm $0.07 \\  
30 Doradus    & 0.76$\pm $0.03 & 0.95$\pm $0.04 & 0.60$\pm $0.04& 0.63$\pm $0.06 \\  
MC North      & 0.60$\pm $0.03 & 0.87$\pm $0.02 & 0.47$\pm $0.04& 0.45$\pm $0.07 \\  
MC South      & 0.76$\pm $0.02 & 0.99$\pm $0.02 & 0.65$\pm $0.02& 0.71$\pm $0.04 \\  
Arcs          & 0.76$\pm $0.02 & 0.97$\pm $0.02 & 0.62$\pm $0.03& 0.71$\pm $0.04 \\  
\hline
Average       & 0.72$\pm $0.03 & 0.94$\pm $0.03 & 0.58$\pm $0.04& 0.62$\pm $0.05 \\  
Average \tablenotemark{b}& 0.68$\pm $0.02& 0.84$\pm $0.02& 0.49$\pm $0.03& 0.57$\pm $0.04 \\  
Average \tablenotemark{c}& 0.78$\pm $0.02& 0.96$\pm $0.02& 0.68$\pm $0.03& 0.71$\pm $0.04 \\  
\hline
HI region     & 0.90$\pm $0.04 & 1.15$\pm $0.04 & 0.83$\pm $0.05& 0.85$\pm $0.06 \\  
LMC bar       & 0.75$\pm $0.01 & 1.05$\pm $0.01 & 0.62$\pm $0.02& 0.67$\pm $0.03 \\  
\hline\hline
\end{tabular}
\tablenotetext{a}{Using $A_{\J}/A_{\Ks}$\,=\,2.96
                          obtained by \citet{Gordon03}
                          for the lines of sight toward
                          the LMC2 supershell near 30 Doradus}
\tablenotetext{b}{With the systematic bias caused by the possible correlated uncertainties of $\J-\Ks$ and
                  $\Ks-\lambda$ corrected.}
\tablenotetext{c}{Using the Galactic value of
                  $A_{\J}/A_{\Ks}$\,=\,2.52 \citep{Rieke85}}
\end{center}
\normalsize
\end{table}

\begin{table}  
\small
\begin{center}
\caption{%
         \label{tab:ice}
         Parameters for the four ice bands of H$_2$O, CO and
         CO$_2$ which may contribute to the mid-IR extinction
         in the Spitzer/IRAC bands
         }
\vspace{0.2in}
\begin{tabular}{lccccccr}
\hline\hline
Band & $\lambda_j$ & Ice Species & FWHM\tablenotemark{a}
        & Band Strength\tablenotemark{b} & Abundance\tablenotemark{c}
        & Enhancement\tablenotemark{d} & Enhancement\tablenotemark{d}\\
$j$     & ($\mu$m) & ${\rm X}_j$ & $\gamma_j$ (cm$^{-1}$)
        & $A_j$ (cm\,mol$^{-1}$)  & ${\rm X}_j/{\rm H_2O}$
        & $E_j$ ($\Rv=3.1$) & $E_j$ ($\Rv=5.5$)      \\
\hline
1 & 3.05 & H$_{2}$O & 335  & $2.0\times10^{-16}$  & 1     & 1.0 & 1.0\\
2 & 4.27 & CO$_{2}$ & 18   & $7.6\times10^{-17}$  & 0.21  & 3.5 & 6.5\\
3 & 4.67 & CO       & 9.7  & $1.1\times10^{-17}$  & 0.25  & 3.5 & 6.5\\
4 & 6.02 & H$_{2}$O & 160  & $8.4\times10^{-18}$  & 1     & 1.0 & 1.0\\
\hline
\end{tabular}
\tablenotetext{a}{\citet{Gibb04}}
\tablenotetext{b}{\citet{Gerakines95}}
\tablenotetext{c}{\citet{Whittet03}}
\tablenotetext{d}{$E_j$ is the ``enhancement'' factor
                  in the sense that in order for CO and
                  CO$_2$ ices to account for the [4.5]
                  excess extinction, the CO and CO$_2$
                  abunadnces need to be ``enhanced''
                  by a factor of $E_j$ (i.e., $E_{\rm CO}$
                  and $E_{\rm CO_2}$) relative to their
                  abundances in typical dense clouds.
                  }
\end{center}
\normalsize
\end{table}


\begin{deluxetable}{cc|cccccc|rrrrrr}
\rotate
\tabletypesize{\scriptsize}
\tablewidth{21cm}
\tablecaption{Results from Monte-Carlo Simulations\label{tab:montecarlo}}
\startdata
\hline\hline
   $N$\tablenotemark{a}
   & $\left(A^{\rm mc}_{\J}\right)_{\rm max}$\tablenotemark{b}
           &$\frac{E(\J-\HH)}{E(\HH-\Ks)}$   & $\frac{E(\J-\Ks)}{E(\HH-\Ks)}$
           &$\frac{E(\Ks-[3.6])}{E(\J-\Ks)}$ & $\frac{E(\Ks-[4.5])}{E(\J-\Ks)}$  & $\frac{E(\Ks-[5.8])}{E(\J-\Ks)}$  & $\frac{E(\Ks-[8.0])}{E(\J-\Ks)}$
           &$\frac{A_{\J}}{A_{\Ks}}$        &$\frac{A_{\HH}}{A_{\Ks}}$  & $\frac{A_{[3.6]}}{A_{\Ks}}$      & $\frac{A_{[4.5]}}{A_{\Ks}}$       & $\frac{A_{[5.8]}}{A_{\Ks}}$       &
           $\frac{A_{[8.0]}}{A_{\Ks}}$     \\
\hline\hline
\multicolumn{2}{c|}{$\Rv=3.1$\tablenotemark{c}}& 1.693& 2.693& 0.404& 0.498& 0.556& 0.539& 2.494 & 1.554& 0.396& 0.256& 0.169& 0.195 \\
 1000  & 0.5   & 1.240& 1.936& 0.275& 0.352& 0.394& 0.415& 2.494& 1.749& 0.589& 0.475& 0.412& 0.381 \\
 1000  & 1.0   & 1.541& 2.407& 0.367& 0.456& 0.510& 0.509& 2.494& 1.601& 0.452& 0.319& 0.238& 0.240 \\
 1000  & 1.5   & 1.620& 2.549& 0.387& 0.478& 0.534& 0.526& 2.494& 1.568& 0.422& 0.285& 0.201& 0.214 \\
 1000  & 5.0   & 1.684& 2.675& 0.402& 0.496& 0.554& 0.538& 2.494& 1.542& 0.399& 0.259& 0.172& 0.196 \\
 1000  & 10.0  & 1.689& 2.687& 0.403& 0.497& 0.555& 0.539& 2.494& 1.539& 0.397& 0.257& 0.170& 0.195 \\
 10000 & 1.5   & 1.612& 2.545& 0.385& 0.479& 0.533& 0.520& 2.494& 1.569& 0.425& 0.284& 0.204& 0.222 \\
 100000& 1.5   & 1.620& 2.553& 0.387& 0.479& 0.536& 0.520& 2.494& 1.568& 0.422& 0.284& 0.199& 0.223 \\
\hline\hline
\multicolumn{2}{c|}{$\Rv=5.5$\tablenotemark{d}} & 2.222& 3.222& 0.276& 0.352& 0.414& 0.421& 2.449 &1.450& 0.600& 0.490& 0.400& 0.389 \\
 1000  & 0.5   & 1.317& 2.050& 0.161& 0.225& 0.271& 0.312& 2.449& 1.707& 0.766& 0.675& 0.608& 0.548 \\
 1000  & 1.0   & 1.884& 2.729& 0.243& 0.314& 0.374& 0.395& 2.449& 1.531& 0.648& 0.545& 0.458& 0.427 \\
 1000  & 1.5   & 2.058& 2.969& 0.260& 0.334& 0.395& 0.411& 2.449& 1.488& 0.623& 0.516& 0.427& 0.405 \\
 1000  & 5.0   & 2.204& 3.193& 0.274& 0.350& 0.412& 0.422& 2.449& 1.454& 0.602& 0.493& 0.403& 0.389 \\
 1000  & 10.0  & 2.214& 3.211& 0.276& 0.351& 0.414& 0.422& 2.449& 1.451& 0.601& 0.491& 0.401& 0.389 \\
 10000 & 1.5   & 2.051& 2.965& 0.259& 0.335& 0.393& 0.405& 2.449& 1.489& 0.625& 0.515& 0.430& 0.413 \\
 100000& 1.5   & 2.060& 2.974& 0.261& 0.335& 0.396& 0.405& 2.449& 1.455& 0.622& 0.515& 0.426& 0.414 \\
\hline
\enddata
\tablenotetext{a}{Number of stars employed in the simulation}
\tablenotetext{b}{The maximum amount of J-band extinction
                          employed to obscure the sources}
\tablenotetext{c}{The MW model extinction curve
                           of $\Rv=3.1$ \citep{Weingartner01}}
\tablenotetext{d}{The MW model extinction curve
                           of $\Rv=5.5$ \citep{Weingartner01}}
\end{deluxetable}

\begin{thebibliography}{69}
\expandafter\ifx\csname natexlab\endcsname\relax\def\natexlab#1{#1}\fi

\bibitem[{{Alves} {et~al.}(2002){Alves}, {Rejkuba}, {Minniti}, \&
  {Cook}}]{Alves02}
{Alves}, D.~R., {Rejkuba}, M., {Minniti}, D., \& {Cook}, K.~H. 2002, \apjl,
  573, L51

\bibitem[{{Bernat}(1981)}]{Bernat81}
{Bernat}, A.~P. 1981, \apj, 246, 184

\bibitem[{{Bessell} {et~al.}(1991){Bessell}, {Brett}, {Scholz}, \&
  {Wood}}]{Bessell91}
{Bessell}, M.~S., {Brett}, J.~M., {Scholz}, M., \& {Wood}, P.~R. 1991, \aaps,
  89, 335

\bibitem[{{Boogert} {et~al.}(2011){Boogert}, {Huard}, {Cook}, {Chiar}, {Knez},
  {Decin}, {Blake}, {Tielens}, \& {van Dishoeck}}]{Boogert11}
{Boogert}, A.~C.~A., {et~al.} 2011, \apj, 729, 92

\bibitem[{{Bressan} {et~al.}(2012){Bressan}, {Marigo}, {Girardi}, {Salasnich},
  {Dal Cero}, {Rubele}, \& {Nanni}}]{Bressan12}
{Bressan}, A., {Marigo}, P., {Girardi}, L., {Salasnich}, B., {Dal Cero}, C.,
  {Rubele}, S., \& {Nanni}, A. 2012, \mnras, 427, 127

\bibitem[{{Cardelli} {et~al.}(1989){Cardelli}, {Clayton}, \&
  {Mathis}}]{Cardelli89}
{Cardelli}, J.~A., {Clayton}, G.~C., \& {Mathis}, J.~S. 1989, \apj, 345, 245

\bibitem[{{Cioni} {et~al.}(2000){Cioni}, {van der Marel}, {Loup}, \&
  {Habing}}]{Cioni00}
{Cioni}, M.-R.~L., {van der Marel}, R.~P., {Loup}, C., \& {Habing}, H.~J. 2000,
  \aap, 359, 601

\bibitem[{{Clayton} \& {Martin}(1985)}]{Clayton85}
{Clayton}, G.~C., \& {Martin}, P.~G. 1985, \apj, 288, 558

\bibitem[{{Cutri} \& {2MASS Team}(2004)}]{Cutri04}
{Cutri}, R.~M., \& {2MASS Team}. 2004, in Bulletin of the American Astronomical
  Society, Vol.~36, American Astronomical Society Meeting Abstracts, 1487

\bibitem[{{Cutri} {et~al.}(2003){Cutri}, {Skrutskie}, {van Dyk}, {Beichman},
  {Carpenter}, {Chester}, {Cambresy}, {Evans}, {Fowler}, {Gizis}, {Howard},
  {Huchra}, {Jarrett}, {Kopan}, {Kirkpatrick}, {Light}, {Marsh}, {McCallon},
  {Schneider}, {Stiening}, {Sykes}, {Weinberg}, {Wheaton}, {Wheelock}, \&
  {Zacarias}}]{Cutri03}
{Cutri}, R.~M., {et~al.} 2003, {2MASS All Sky Catalog of point sources.}

\bibitem[{{Dobashi} {et~al.}(2008){Dobashi}, {Bernard}, {Hughes}, {Paradis},
  {Reach}, \& {Kawamura}}]{Dobashi08}
{Dobashi}, K., {Bernard}, J.-P., {Hughes}, A., {Paradis}, D., {Reach}, W.~T.,
  \& {Kawamura}, A. 2008, \aap, 484, 205

\bibitem[{{Draine}(1989)}]{Draine89}
{Draine}, B.~T. 1989, in ESA Special Publication, Vol. 290, Infrared
  Spectroscopy in Astronomy, ed. E.~{B{\"o}hm-Vitense}, 93--98

\bibitem[{{Draine}(2003)}]{Draine03}
{Draine}, B.~T. 2003, \araa, 41, 241

\bibitem[{{Fitzpatrick}(1985)}]{Fitzpatrick85}
{Fitzpatrick}, E.~L. 1985, \apj, 299, 219

\bibitem[{{Fitzpatrick}(1986)}]{Fitzpatrick86}
---. 1986, \aj, 92, 1068

\bibitem[{{Fitzpatrick} \& {Massa}(2009)}]{Fitzpatrick09}
{Fitzpatrick}, E.~L., \& {Massa}, D. 2009, \apj, 699, 1209

\bibitem[{{Flaherty} {et~al.}(2007){Flaherty}, {Pipher}, {Megeath}, {Winston},
  {Gutermuth}, {Muzerolle}, {Allen}, \& {Fazio}}]{Flaherty07}
{Flaherty}, K.~M., {Pipher}, J.~L., {Megeath}, S.~T., {Winston}, E.~M.,
  {Gutermuth}, R.~A., {Muzerolle}, J., {Allen}, L.~E., \& {Fazio}, G.~G. 2007,
  \apj, 663, 1069

\bibitem[{{Fukui} {et~al.}(2008){Fukui}, {Kawamura}, {Minamidani}, {Mizuno},
  {Kanai}, {Mizuno}, {Onishi}, {Yonekura}, {Mizuno}, {Ogawa}, \&
  {Rubio}}]{Fukui08}
{Fukui}, Y., {et~al.} 2008, \apjs, 178, 56

\bibitem[{{Gao} {et~al.}(2009){Gao}, {Jiang}, \& {Li}}]{Gao09}
{Gao}, J., {Jiang}, B.~W., \& {Li}, A. 2009, \apj, 707, 89

\bibitem[{{Garrod} \& {Pauly}(2011)}]{Garrod11}
{Garrod}, R.~T., \& {Pauly}, T. 2011, \apj, 735, 15

\bibitem[{{Gerakines} {et~al.}(1995){Gerakines}, {Schutte}, {Greenberg}, \&
  {van Dishoeck}}]{Gerakines95}
{Gerakines}, P.~A., {Schutte}, W.~A., {Greenberg}, J.~M., \& {van Dishoeck},
  E.~F. 1995, \aap, 296, 810

\bibitem[{{Gibb} {et~al.}(2004){Gibb}, {Whittet}, {Boogert}, \&
  {Tielens}}]{Gibb04}
{Gibb}, E.~L., {Whittet}, D.~C.~B., {Boogert}, A.~C.~A., \& {Tielens},
  A.~G.~G.~M. 2004, \apjs, 151, 35

\bibitem[{{Gordon} \& {Clayton}(1998)}]{Gordon98}
{Gordon}, K.~D., \& {Clayton}, G.~C. 1998, \apj, 500, 816

\bibitem[{{Gordon} {et~al.}(2003){Gordon}, {Clayton}, {Misselt}, {Landolt}, \&
  {Wolff}}]{Gordon03}
{Gordon}, K.~D., {Clayton}, G.~C., {Misselt}, K.~A., {Landolt}, A.~U., \&
  {Wolff}, M.~J. 2003, \apj, 594, 279

\bibitem[{{Imara} \& {Blitz}(2007)}]{Imara07}
{Imara}, N., \& {Blitz}, L. 2007, \apj, 662, 969

\bibitem[{{Indebetouw} {et~al.}(2005){Indebetouw}, {Mathis}, {Babler}, {Meade},
  {Watson}, {Whitney}, {Wolff}, {Wolfire}, {Cohen}, {Bania}, {Benjamin},
  {Clemens}, {Dickey}, {Jackson}, {Kobulnicky}, {Marston}, {Mercer},
  {Stauffer}, {Stolovy}, \& {Churchwell}}]{Indebetouw05}
{Indebetouw}, R., {et~al.} 2005, \apj, 619, 931

\bibitem[{{Ioppolo} {et~al.}(2011){Ioppolo}, {van Boheemen}, {Cuppen}, {van
  Dishoeck}, \& {Linnartz}}]{Ioppolo11}
{Ioppolo}, S., {van Boheemen}, Y., {Cuppen}, H.~M., {van Dishoeck}, E.~F., \&
  {Linnartz}, H. 2011, \mnras, 413, 2281

\bibitem[{{Israel} {et~al.}(1986){Israel}, {de Graauw}, {van de Stadt}, \& {de
  Vries}}]{Israel86}
{Israel}, F.~P., {de Graauw}, T., {van de Stadt}, H., \& {de Vries}, C.~P.
  1986, \apj, 303, 186

\bibitem[{{Jiang} {et~al.}(2006){Jiang}, {Gao}, {Omont}, {Schuller}, \&
  {Simon}}]{Jiang06}
{Jiang}, B.~W., {Gao}, J., {Omont}, A., {Schuller}, F., \& {Simon}, G. 2006,
  \aap, 446, 551

\bibitem[{{Jiang} {et~al.}(2003){Jiang}, {Omont}, {Ganesh}, {Simon}, \&
  {Schuller}}]{Jiang03}
{Jiang}, B.~W., {Omont}, A., {Ganesh}, S., {Simon}, G., \& {Schuller}, F. 2003,
  \aap, 400, 903

\bibitem[{{Kerber} {et~al.}(2009){Kerber}, {Girardi}, {Rubele}, \&
  {Cioni}}]{Kerber09}
{Kerber}, L.~O., {Girardi}, L., {Rubele}, S., \& {Cioni}, M.-R. 2009, \aap,
  499, 697

\bibitem[{{Koornneef}(1982)}]{Koornneef82}
{Koornneef}, J. 1982, \aap, 107, 247

\bibitem[{{Koornneef} \& {Code}(1981)}]{Koornneef81}
{Koornneef}, J., \& {Code}, A.~D. 1981, \apj, 247, 860

\bibitem[{{Lacy} {et~al.}(1984){Lacy}, {Baas}, {Allamandola}, {van de Bult},
  {Persson}, {McGregor}, {Lonsdale}, \& {Geballe}}]{Lacy84}
{Lacy}, J.~H., {Baas}, F., {Allamandola}, L.~J., {van de Bult}, C.~E.~P.,
  {Persson}, S.~E., {McGregor}, P.~J., {Lonsdale}, C.~J., \& {Geballe}, T.~R.
  1984, \apj, 276, 533

\bibitem[{{Lada} {et~al.}(1994){Lada}, {Lada}, {Clemens}, \& {Bally}}]{Lada94}
{Lada}, C.~J., {Lada}, E.~A., {Clemens}, D.~P., \& {Bally}, J. 1994, \apj, 429,
  694

\bibitem[{{Lequeux} {et~al.}(1982){Lequeux}, {Maurice}, {Prevot-Burnichon},
  {Prevot}, \& {Rocca-Volmerange}}]{Lequeux82}
{Lequeux}, J., {Maurice}, E., {Prevot-Burnichon}, M.-L., {Prevot}, L., \&
  {Rocca-Volmerange}, B. 1982, \aap, 113, L15

\bibitem[{{Li} \& {Mann}(2012)}]{Li12}
{Li}, A., \& {Mann}, I. 2012, in Astrophysics and Space Science Library, Vol.
  385, Astrophysics and Space Science Library, ed. I.~{Mann},
  N.~{Meyer-Vernet}, \& A.~{Czechowski}, 5

\bibitem[{{Lombardi} \& {Alves}(2001)}]{Lombardi01}
{Lombardi}, M., \& {Alves}, J. 2001, \aap, 377, 1023

\bibitem[{{Meixner} {et~al.}(2006){Meixner}, {Gordon}, {Indebetouw}, {Hora},
  {Whitney}, {Blum}, {Reach}, {Bernard}, {Meade}, {Babler}, {Engelbracht},
  {For}, {Misselt}, {Vijh}, {Leitherer}, {Cohen}, {Churchwell}, {Boulanger},
  {Frogel}, {Fukui}, {Gallagher}, {Gorjian}, {Harris}, {Kelly}, {Kawamura},
  {Kim}, {Latter}, {Madden}, {Markwick-Kemper}, {Mizuno}, {Mizuno}, {Mould},
  {Nota}, {Oey}, {Olsen}, {Onishi}, {Paladini}, {Panagia}, {Perez-Gonzalez},
  {Shibai}, {Sato}, {Smith}, {Staveley-Smith}, {Tielens}, {Ueta}, {van Dyk},
  {Volk}, {Werner}, \& {Zaritsky}}]{Meixner06}
{Meixner}, M., {et~al.} 2006, \aj, 132, 2268

\bibitem[{{Mennella} {et~al.}(2004){Mennella}, {Palumbo}, \&
  {Baratta}}]{Mennella04}
{Mennella}, V., {Palumbo}, M.~E., \& {Baratta}, G.~A. 2004, \apj, 615, 1073

\bibitem[{{Misselt} {et~al.}(1999){Misselt}, {Clayton}, \&
  {Gordon}}]{Misselt99}
{Misselt}, K.~A., {Clayton}, G.~C., \& {Gordon}, K.~D. 1999, \apj, 515, 128

\bibitem[{{Mucciarelli} {et~al.}(2006){Mucciarelli}, {Origlia}, {Ferraro},
  {Maraston}, \& {Testa}}]{Mucciarelli06}
{Mucciarelli}, A., {Origlia}, L., {Ferraro}, F.~R., {Maraston}, C., \& {Testa},
  V. 2006, \apj, 646, 939

\bibitem[{{Nandy} {et~al.}(1981){Nandy}, {Morgan}, {Willis}, {Wilson}, \&
  {Gondhalekar}}]{Nandy81}
{Nandy}, K., {Morgan}, D.~H., {Willis}, A.~J., {Wilson}, R., \& {Gondhalekar},
  P.~M. 1981, \mnras, 196, 955

\bibitem[{{Nikolaev} \& {Weinberg}(2000)}]{Nikolaev00}
{Nikolaev}, S., \& {Weinberg}, M.~D. 2000, \apj, 542, 804

\bibitem[{{Nishiyama} {et~al.}(2009){Nishiyama}, {Tamura}, {Hatano}, {Kato},
  {Tanab{\'e}}, {Sugitani}, \& {Nagata}}]{Nishiyama09}
{Nishiyama}, S., {Tamura}, M., {Hatano}, H., {Kato}, D., {Tanab{\'e}}, T.,
  {Sugitani}, K., \& {Nagata}, T. 2009, \apj, 696, 1407

\bibitem[{{Noble} {et~al.}(2011){Noble}, {Dulieu}, {Congiu}, \&
  {Fraser}}]{Noble11}
{Noble}, J.~A., {Dulieu}, F., {Congiu}, E., \& {Fraser}, H.~J. 2011, \apj, 735,
  121

\bibitem[{{Nummelin} {et~al.}(2001){Nummelin}, {Whittet}, {Gibb}, {Gerakines},
  \& {Chiar}}]{Nummelin01}
{Nummelin}, A., {Whittet}, D.~C.~B., {Gibb}, E.~L., {Gerakines}, P.~A., \&
  {Chiar}, J.~E. 2001, \apj, 558, 185

\bibitem[{{Oba} {et~al.}(2010){Oba}, {Watanabe}, {Kouchi}, {Hama}, \&
  {Pirronello}}]{Oba10}
{Oba}, Y., {Watanabe}, N., {Kouchi}, A., {Hama}, T., \& {Pirronello}, V. 2010,
  \apjl, 712, L174

\bibitem[{{Oestreicher} {et~al.}(1995){Oestreicher}, {Gochermann}, \&
  {Schmidt-Kaler}}]{Oestreicher95}
{Oestreicher}, M.~O., {Gochermann}, J., \& {Schmidt-Kaler}, T. 1995, \aaps,
  112, 495

\bibitem[{{Pendleton} {et~al.}(1999){Pendleton}, {Tielens}, {Tokunaga}, \&
  {Bernstein}}]{Pendleton99}
{Pendleton}, Y.~J., {Tielens}, A.~G.~G.~M., {Tokunaga}, A.~T., \& {Bernstein},
  M.~P. 1999, \apj, 513, 294

\bibitem[{{Prevot} {et~al.}(1984){Prevot}, {Lequeux}, {Prevot}, {Maurice}, \&
  {Rocca-Volmerange}}]{Prevot84}
{Prevot}, M.~L., {Lequeux}, J., {Prevot}, L., {Maurice}, E., \&
  {Rocca-Volmerange}, B. 1984, \aap, 132, 389

\bibitem[{{Rieke} \& {Lebofsky}(1985)}]{Rieke85}
{Rieke}, G.~H., \& {Lebofsky}, M.~J. 1985, \apj, 288, 618

\bibitem[{{Ruffle} \& {Herbst}(2001)}]{Ruffle01}
{Ruffle}, D.~P., \& {Herbst}, E. 2001, \mnras, 324, 1054

\bibitem[{{Russell} \& {Dopita}(1992)}]{Russell92}
{Russell}, S.~C., \& {Dopita}, M.~A. 1992, \apj, 384, 508

\bibitem[{{Sakai} {et~al.}(2000){Sakai}, {Zaritsky}, \& {Kennicutt}}]{Sakai00}
{Sakai}, S., {Zaritsky}, D., \& {Kennicutt}, Jr., R.~C. 2000, \aj, 119, 1197

\bibitem[{{Sakon} {et~al.}(2006){Sakon}, {Onaka}, {Kaneda}, {Tokura}, {Takagi},
  {Tajiri}, {Takahashi}, {Kato}, {Onishi}, {Kawamura}, \& {Fukui}}]{Sakon06}
{Sakon}, I., {et~al.} 2006, \apj, 651, 174

\bibitem[{{Salaris} \& {Girardi}(2005)}]{Salaris05}
{Salaris}, M., \& {Girardi}, L. 2005, \mnras, 357, 669

\bibitem[{{Schlegel} {et~al.}(1998){Schlegel}, {Finkbeiner}, \&
  {Davis}}]{Schlegel98}
{Schlegel}, D.~J., {Finkbeiner}, D.~P., \& {Davis}, M. 1998, \apj, 500, 525

\bibitem[{{Schutte} \& {Greenberg}(1997)}]{Schutte97}
{Schutte}, W.~A., \& {Greenberg}, J.~M. 1997, \aap, 317, L43

\bibitem[{{Schwering} \& {Israel}(1991)}]{Schwering91}
{Schwering}, P.~B.~W., \& {Israel}, F.~P. 1991, \aap, 246, 231

\bibitem[{{Shimonishi} {et~al.}(2008){Shimonishi}, {Onaka}, {Kato}, {Sakon},
  {Ita}, {Kawamura}, \& {Kaneda}}]{Shimonishi08}
{Shimonishi}, T., {Onaka}, T., {Kato}, D., {Sakon}, I., {Ita}, Y., {Kawamura},
  A., \& {Kaneda}, H. 2008, \apjl, 686, L99

\bibitem[{{Staveley-Smith} {et~al.}(2003){Staveley-Smith}, {Kim}, {Calabretta},
  {Haynes}, \& {Kesteven}}]{Staveley-Smith03}
{Staveley-Smith}, L., {Kim}, S., {Calabretta}, M.~R., {Haynes}, R.~F., \&
  {Kesteven}, M.~J. 2003, \mnras, 339, 87

\bibitem[{{Tatton} {et~al.}(2013){Tatton}, {van Loon}, {Cioni}, {Clementini},
  {Emerson}, {Girardi}, {de Grijs}, {Groenewegen}, {Gullieuszik}, {Ivanov},
  {Moretti}, {Ripepi}, \& {Rubele}}]{Tatton13}
{Tatton}, B.~L., {et~al.} 2013, ArXiv e-prints

\bibitem[{{Weingartner} \& {Draine}(2001)}]{Weingartner01}
{Weingartner}, J.~C., \& {Draine}, B.~T. 2001, \apj, 548, 296

\bibitem[{{Whittet}(2003)}]{Whittet03}
{Whittet}, D.~C.~B., ed. 2003, {Dust in the galactic environment}, ed. D.~C.~B.
  {Whittet}

\bibitem[{{Whittet} {et~al.}(2001{\natexlab{a}}){Whittet}, {Gerakines},
  {Hough}, \& {Shenoy}}]{Whittet01a}
{Whittet}, D.~C.~B., {Gerakines}, P.~A., {Hough}, J.~H., \& {Shenoy}, S.~S.
  2001{\natexlab{a}}, \apj, 547, 872

\bibitem[{{Whittet} {et~al.}(2001{\natexlab{b}}){Whittet}, {Pendleton}, {Gibb},
  {Boogert}, {Chiar}, \& {Nummelin}}]{Whittet01b}
{Whittet}, D.~C.~B., {Pendleton}, Y.~J., {Gibb}, E.~L., {Boogert}, A.~C.~A.,
  {Chiar}, J.~E., \& {Nummelin}, A. 2001{\natexlab{b}}, \apj, 550, 793

\bibitem[{{Whittet} {et~al.}(2007){Whittet}, {Shenoy}, {Bergin}, {Chiar},
  {Gerakines}, {Gibb}, {Melnick}, \& {Neufeld}}]{Whittet07}
{Whittet}, D.~C.~B., {Shenoy}, S.~S., {Bergin}, E.~A., {Chiar}, J.~E.,
  {Gerakines}, P.~A., {Gibb}, E.~L., {Melnick}, G.~J., \& {Neufeld}, D.~A.
  2007, \apj, 655, 332

\bibitem[{{Zaritsky} {et~al.}(2004){Zaritsky}, {Harris}, {Thompson}, \&
  {Grebel}}]{Zaritsky04}
{Zaritsky}, D., {Harris}, J., {Thompson}, I.~B., \& {Grebel}, E.~K. 2004, \aj,
  128, 1606

\end{thebibliography}
\end{document}